\documentclass[journal,transmag]{IEEEtran}
\usepackage{amsmath}
\usepackage{amsfonts}
\usepackage{url}
\usepackage{amssymb}
\usepackage[]{graphicx}
\usepackage{hyperref}
\ifCLASSINFOpdf
\else
\fi
\begin{document}
\title{Efficient calculation of the self magnetic field, self-force, and self-inductance for electromagnetic coils}


\author{\IEEEauthorblockN{Siena Hurwitz\IEEEauthorrefmark{1},
Matt Landreman\IEEEauthorrefmark{1},
Thomas M. Antonsen, Jr\IEEEauthorrefmark{1},
}
\IEEEauthorblockA{\IEEEauthorrefmark{1}Institute for Research in Electronics and Applied Physics, University of Maryland, College Park, MD 21043 USA}% <-this % stops an unwanted space
% \thanks{Manuscript received December 1, 2012; revised August 26, 2015. 
% Corresponding author: M. Shell (email: http://www.michaelshell.org/contact.html).}
}

%{Shell \MakeLowercase{\textit{et al.}}: Bare Demo of IEEEtran.cls for IEEE Transactions on Magnetics Journals}



% for Transactions on Magnetics papers, we must declare the abstract and
% index terms PRIOR to the title within the \IEEEtitleabstractindextext
% IEEEtran command as these need to go into the title area created by
\IEEEtitleabstractindextext{%
\begin{abstract}
The design of electromagnetic coils may require evaluation of several quantities that are challenging to compute numerically. These quantities include Lorentz forces, which may be a limiting factor due to stresses; the internal magnetic field, which is relevant for determining stress as well as a superconducting coil's proximity to its quench limit; and the inductance, which determines stored magnetic energy and dynamics. When computing the effect on one coil due to the current in another, these quantities can often be approximated quickly by treating the coils as infinitesimally thin. When computing the effect on a coil due to its own current (e.g., self-force or self-inductance), evaluation is difficult due to the presence of a singularity; coils cannot be treated as infinitesimally thin as each quantity diverges at zero conductor width. Here, we present novel and well-behaved methods for evaluating these quantities using non-singular integral formulae of reduced dimensions. These formulae are determined rigorously by dividing the domain of integration of the magnetic vector potential into two regions, exploiting appropriate approximations in each region, and expanding in high aspect ratio. Our formulae show good agreement to full finite-thickness calculations even at low aspect ratio, both analytically for a torus and numerically for a non-planar coil of a stellarator fusion device, the Helically Symmetric eXperiment (HSX). Because the integrands of these formulae develop fine structure as the minor radius becomes infinitely thin, we also develop a method of evaluating the self-force and self-inductance with even greater efficiency by integrating this sharp feature analytically. We demonstrate with this method that the self-force can be accurately computed for the HSX coil with as few as 12 grid points.
\end{abstract}

}
\maketitle
\IEEEdisplaynontitleabstractindextext
\IEEEpeerreviewmaketitle

\section{Introduction}
\IEEEPARstart{T}{he} self-field, self-force, and self-inductance are characteristic quantities of conducting coils that have both theoretical and practical importance. One approach to solve for these quantities is to use finite element techniques. For coils with complicated geometries, this approach can lead to high-dimensional numerical grids and a computationally intensive system of equations. Alternatively, one can use equations such as the Biot-Savart Law with a finite-thickness coil representation to solve for these quantities, though numerical evaluation is generally difficult due to the high dimensionality of the integrals and an integrable singularity. Notably, these finite-thickness equations can be reduced to one- and two-dimensional integrals when considering effects between separate coils (such as the mutual inductance) by treating each coil as an infinitely thin filament. These cases also do not suffer from singularities. However, one cannot  perform a similar filamentary treatment for the self-force, internal magnetic field, and self-inductance as these quantities diverge when the conductor approaches zero thickness. Instead, one must consider the three-dimensional structure of the conductor and integrate over the singularity to obtain a correct result. To expedite these calculations, in the present paper we present novel well-behaved methods for evaluating these quantities using non-singular integral formulae of reduced dimensions. 

Some prior research has been done to simplify calculations of the self-force, self-inductance, and internal magnetic field. 
In \cite{sackett1975calculation,urankar1982vector,lion2021general}, the internal magnetic field of conductors was computed in a limited geometry: rectangular cross-section and a center-line made of straight segments and circular arcs. 
In \cite{garren1994lorentz}, an energy-based derivation of the self-force was presented to obtain a one-dimensional integral approximation.
A simpler integral form of the self-inductance was found in \cite{dengler2012self} using a methodology not dissimilar to our own.
In these last two works, a section of the domain was removed from the final integral,  whereas in our approach the integral will be performed over the full length of the filamentary coil.
Integrating over the full periodic filament is advantageous as spectral convergence can be achieved with a uniform quadrature grid, and data on fixed quadrature points can be reused to evaluate the self-field at other points along the coil.
Ref \cite{robin2022minimization} examined a current sheet rather than discrete coils, and the self-force of a current sheet was computed in terms of several surface integrals.

In this paper, we rigorously derive filamentary integral models for the self-inductance (\ref{91}), magnetic field inside and near the coil (\ref{102})-(\ref{63}), and self-force (\ref{103}), for thin coils with circular cross-section and uniformly-distributed current density. These integrals are of reduced dimensionality and generally resemble the standard expressions in the limit in which the minor radius is zero, though they have an added regularization term in the denominator that depends on the square of the minor radius. This regularization term therefore prevents the integrand from diverging when the source and evaluation points are the same, and the scale of the minor radius determines the domain in which the regularization term is non-negligible. A circular cross-section is not representative of many real-world coils so our formulae may therefore be less accurate in these cases. We present the circular cross-section case here as the algebraic complexity is less than for other cross-section shapes, allowing a clearer emphasis on the asymptotic methods. The more complicated case of a rectangular cross-section will be addressed in a subsequent accompanying paper \cite{PaperII}.

\begin{figure}[!htb]
        \center{\includegraphics[width=0.7\columnwidth]{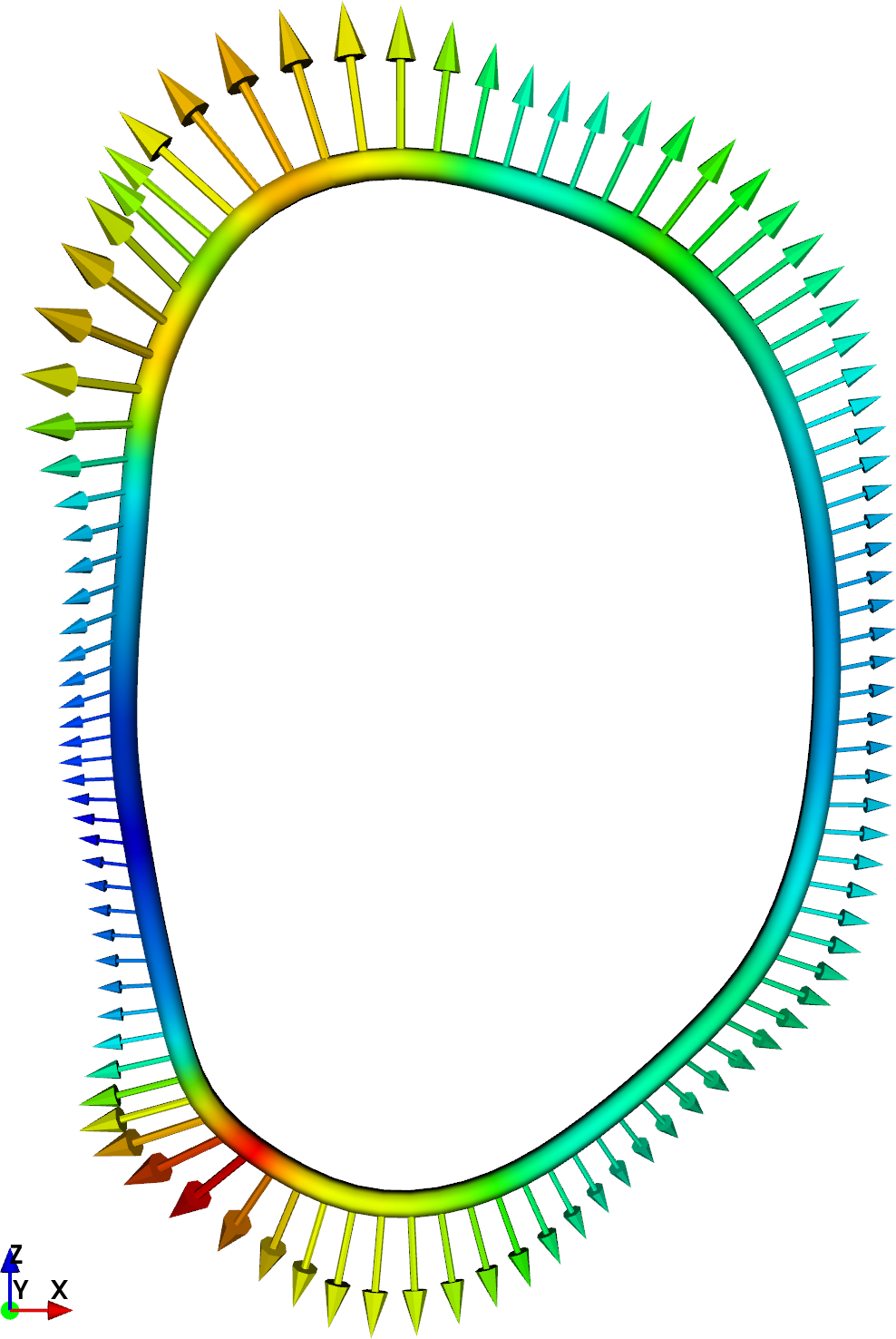}} 
        \caption{\label{fig:3D_force} The figure shows the self-force per unit length across the HSX coil. The direction of the arrows indicates the direction of the force, while the size and color of the arrows indicates the magnitude of the force. }
        \end{figure}

Our proofs, discussed in Section \ref{sec:simplifiedform} and completed in the Appendices, rely on partitioning the domain of the integral for the magnetic vector potential across the length of the coil into a ``near region'' and a ``far region.'' The integral in the near region may be evaluated analytically, while the integral in the far region can be simplified to a 1D integral with no singularity. These results show remarkable numerical agreement with the standard finite thickness calculations, referred to throughout this paper as ``exact'' calculations. We show this correspondence analytically for a circular loop and numerically for a non-planar coil of a stellarator plasma confinement device, the Helically Symmetric eXperiment (HSX) at the University of Wisconsin \cite{HSX}. Figure \ref{fig:3D_force} illustrates the self-force on one of the HSX coils. Because the integrands of our novel formulae become sharp at the evaluation point when the minor radius approaches zero, we also discuss methods of making the numerical evaluations even more efficient in Section \ref{sec:quad} by eliminating this fine-scale structure in the integrals. We demonstrate that quadrature for the self-force can be performed accurately with this method for the HSX coil with a grid size of as small as 12 points. The code used for generating data in this project is available at \url{https://github.com/smhurwitz/Coil_Fields}.

The results of this paper may be especially relevant to the design of future nuclear fusion reactors as higher field strengths will enable better confinement of plasmas \cite{robin2022minimization}. Use of high-temperature superconductors (HTS), which can operate at higher field strengths than low-temperature superconductors and conventional conductors, is plausible in the near future. Currently, several superconductors such as niobium-titanium (NbTi) and niobium-tin (Nb3Sn) are well established for fusion applications, and  HTS materials such as rare earth barium copper oxide (REBCO) are promising \cite{national2021bringing, riva2023development}. REBCO, for example, has already been incorporated into future reactor designs, and the SPARC facility aims to test the material's viability in a fusion context\cite{creely2020overview}. Several start-up companies such as Commonwealth Fusion, Tokamak Energy, Type One Energy, Renaissance Fusion, and Stellarex have also placed an emphasis on HTS.

A consequence of the incorporation of superconducting coils into fusion designs and subsequent use of high-strength magnetic fields is that quantities such as the self-force, self-inductance, and internal magnetic field become more critical to coil design. One emerging limiting factor in coil design is the mechanical stress, described by the self-force density and the internal magnetic fields, which the coils and their support structures experience due to Lorentz forces\cite{lion2021general}. For instance, the European DEMO project faced difficulties in which stresses in the coil casing and conductor jacket exceeded allowable limits and required a redesign of the toroidal field coils.\cite{panin2016approaches}. In another case, the edge magnetic island chain in the W7-X stellarator shifted several centimeters outwards at higher magnetic field strengths due to coil deformations \cite{pedersen2016confirmation}. The self-inductance $L$ of a coil may also be relevant to reactor designs as it determines the stored magnetic energy $U$ through $U=(1/2)LI^2$, where $I$ is the coil current. It may important to assess the magnetic energy as, if a superconductor ever leaves its superconducting state such as due a quench, $U$ represents the maximum amount of energy dissipated resistivity that cooling systems must accommodate.

There exist a number of applications of our work to address these and other problems. The broadest is the fast evaluation of various magnet designs, which can be used for purposes both within and outside of fusion. For stellarators in particular, the simplified formulae can be used to optimize coil shapes to minimize magnetic stress and stored magnetic energy. They may also be used to optimize coil shapes with respect to the superconducting quench limit, though this practice may be less accurate as we assume in this paper that current density in a coil is uniform, an approximation of questionable accuracy for superconductors. Our results may also be relevant to other contexts in which coil optimization is important, such as for magnetic resonance imaging (MRI) \cite{hidalgo2010theory, chen2017electromagnetic}
and particle accelerators \cite{russenschuck2011field}.

\section{Coordinate System\label{sec:frenet}}
\begin{figure}[!htb]
        \center{\includegraphics[width=0.7\columnwidth]
        {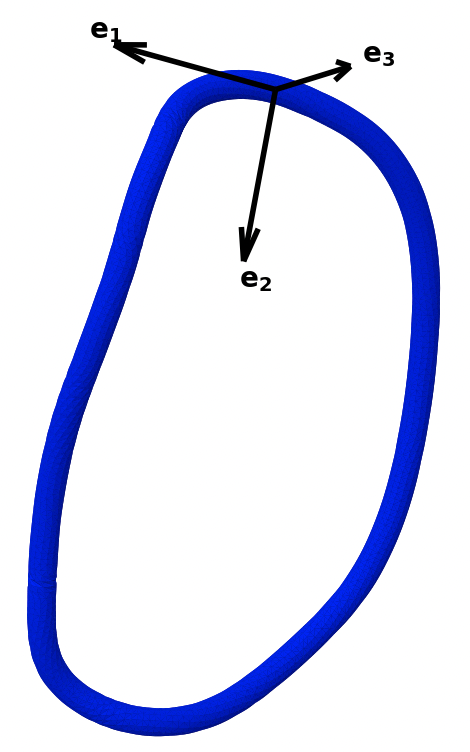}}
        \caption{\label{fig:surface}The orthogonal Frenet-Serret unit vectors are shown on the first HSX coil. The vector $\textbf{e}_1$ points parallel to $\textbf{r}_c$, $\textbf{e}_2$ points in the direction of curvature of $\textbf{r}_c$, and $\textbf{e}_3=\textbf{e}_1\times\textbf{e}_2$.}
        \end{figure}

We describe a coil in terms of a minor radius, $a$, and a center-line, $\textbf{r}_c(\phi)$, where $\phi\in[0,2\pi)$  parameterizes $\textbf{r}_c$. 
We consider the Frenet-Serret frame associated with $\textbf{r}_c$, a right-handed set of orthonormal basis vectors, $\{\textbf{e}_1(\phi),\textbf{e}_2(\phi), \textbf{e}_3(\phi)\}$ \cite{kreyszig_1991}. These three unit vectors are oriented such that $\textbf{e}_1$ lies tangent to $\textbf{r}_c$, $\textbf{e}_2$ lies in the direction of curvature of $\textbf{r}_c$, and $\textbf{e}_3$ lies orthogonal to both $\textbf{e}_1$ and $\textbf{e}_2$. Explicitly,
\begin{subequations}\label{1}
\begin{gather}
\textbf{e}_1=\textbf{r}_c'/|\textbf{r}_c'|\label{1a}\\
\textbf{e}_2=\textbf{e}_1'/|\textbf{e}_1'|\label{1b}\\
\textbf{e}_3=\textbf{e}_1\times\textbf{e}_2\label{1c},
\end{gather}
\end{subequations}
where primes indicate differentiation with respect to $\phi$. The vectors satisfy the Frenet-Serret formulas,
\begin{equation}\label{2}
\frac{d}{d\phi}\begin{pmatrix}
\textbf{e}_1 \\ \textbf{e}_2 \\ \textbf{e}_3\end{pmatrix}=|\textbf{r}_c'|\begin{pmatrix} 0 & {\kappa} & 0\\ -{\kappa} & 0 & {\tau} \\ 0 & -{\tau} & 0
\end{pmatrix}\begin{pmatrix}
\textbf{e}_1 \\ \textbf{e}_2 \\ \textbf{e}_3\end{pmatrix},
\end{equation}
where the curvature and torsion, $\kappa(\phi)$ and $\tau(\phi)$, are given by
\begin{subequations}\label{3}
    \begin{gather}
        {\kappa}=\frac{|\textbf{r}_c'\times\textbf{r}_c''|}{|\textbf{r}_c'|^3}\label{3a}\\
\tau=\frac{(\textbf{r}_c'\times\textbf{r}_c'')\cdot\textbf{r}_c'''}{|\textbf{r}_c'\times\textbf{r}_c''|^2}.\label{3b}
    \end{gather}
\end{subequations}
These vectors are demonstrated in Figure \ref{fig:surface} for the HSX coil. 
All results of this paper behave smoothly as $\kappa\to 0$, and $\tau$ does not appear in the final expressions, so results are applicable even if the coil center-line has straight points or intervals. 

The Frenet basis allows us to construct a  coordinate system defined by the position vector
\begin{equation}\label{4}
\textbf{r}(s,\theta,\phi)=\textbf{r}_c+s\cos\theta\textbf{e}_2+s\sin\theta\textbf{e}_3,
\end{equation}
where $\phi$ behaves as a toroidal angle, $\theta\in[0,2\pi)$ sweeps across the plane orthogonal to $\textbf{r}_c$ and behaves as a poloidal angle, and $s$ is a radius in the same plane. Using (\ref{2})  and (\ref{4}), the Jacobian is
\begin{equation}
    \sqrt{g}=\frac{\partial\textbf{r}}{\partial\phi}\cdot\left(\frac{\partial\textbf{r}}{\partial s}\times\frac{\partial\textbf{r}}{\partial\theta}\right)
=s(1-\kappa s\cos\theta)|\textbf{r}_c'|.
\label{7}
\end{equation}

\section{Simplified Formulae for Thin Coils}\label{sec:simplifiedform}
In this Section, we discuss the simplified integral formulae for the self-inductance, self magnetic field, and self-force density of thin electromagnetic coils. These coils are chosen to carry a uniformly distributed current $I$, where the current density can be written as
\begin{equation}\label{130}
    \textbf{J}=\frac{I}{\pi a^2}\textbf{e}_1.
\end{equation}
It can be shown that this expression satisfies
$\nabla\cdot\textbf{J}=0$. We expand in the small parameter $a/\mathcal{L}$, which behaves as an inverse aspect ratio, where $\mathcal{L}$ represents length scales associated with $\textbf{r}_c$ such as $|\textbf{r}_c'|$,  $\kappa^{-1}$, and $\tau^{-1}$. To obtain a concrete measure of the scale length $\mathcal{L}$, we introduce the quantity $R$, which is defined to be the coil length divided by $2\pi$, and can be conceptualized as an approximate major radius. 

\subsection{Motivations}

\begin{figure}[!htb]
        \center{\includegraphics[width=\columnwidth]
        {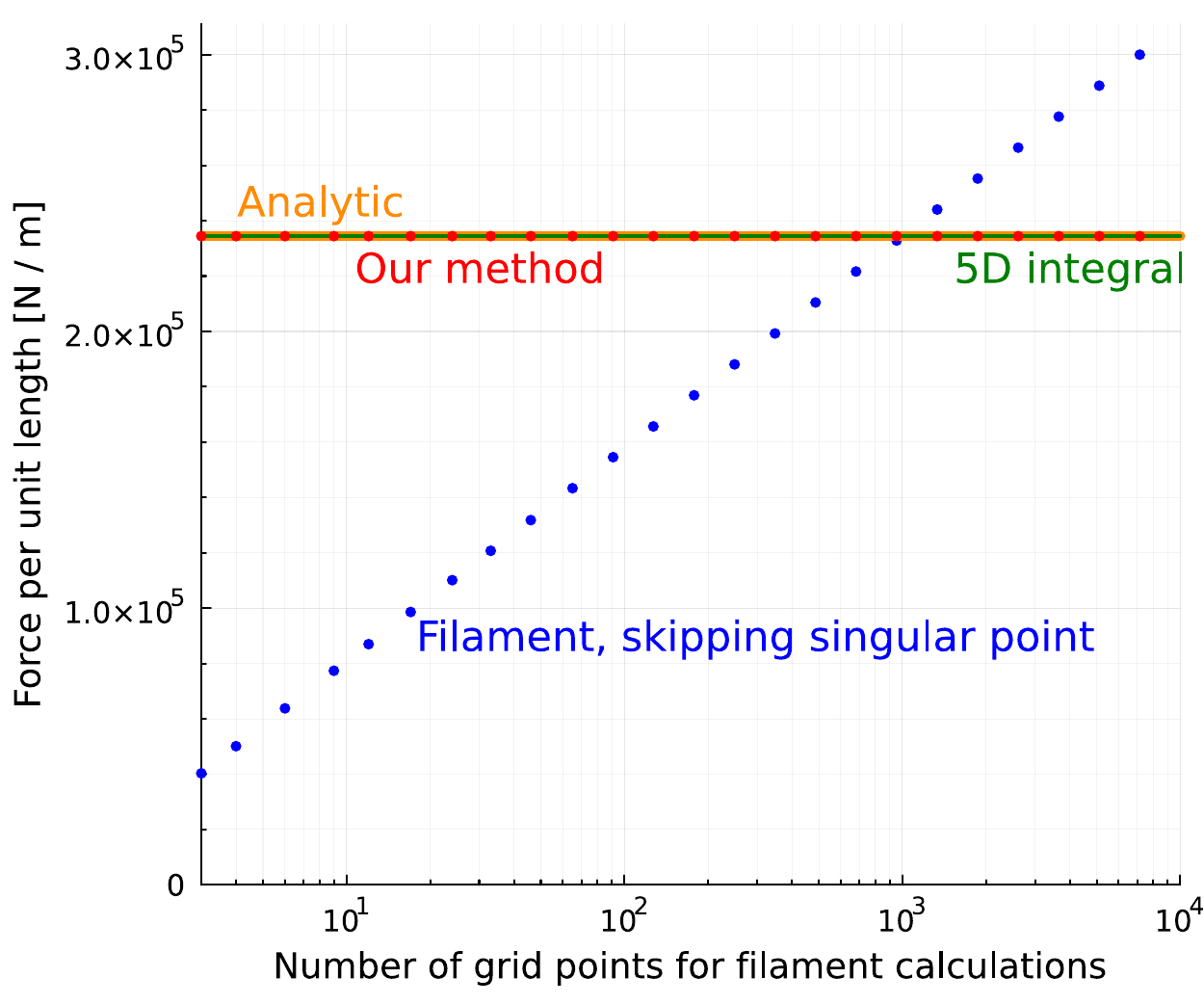}}
        \caption{\label{fig:skipping}Demonstration that dropping the singular grid point at $\tilde{\textbf{r}}_c=\textbf{r}_c$ for a numerical integration of (\ref{127})-(\ref{100}) is an invalid method of determining self-force. This is shown by comparing the results of this method at various grid resolutions to the analytic solution (\ref{105}), a finite-thickness numerical evaluation (\ref{129}), and our method (\ref{103}) and (\ref{108}) for a circular coil. All methods agree with one another except that in which the singularity is dropped. The conductor is circular with $I=10^6$ A and $R/a=300$.}
        \end{figure}

As mentioned in the Introduction, it is essential to consider the three-dimensional structure of a conductor when determining quantities like the self-inductance or self-force; attempts to let $a/\mathcal{L}\rightarrow 0$ and use the one-dimensional form of the Biot-Savart law or two-dimensional form of the self-inductance formula inherently fail. To demonstrate this, we consider the self-force on a circular conductor. When letting $a/\mathcal{L}\rightarrow 0$, one could write the self-force density as 
\begin{equation}\label{127}
    \frac{d\textbf{F}_\textrm{1D}}{dl}=I\textbf{e}_1\times\textbf{B}_\textrm{1D},
\end{equation}
where the magnetic field is given by the one-dimensional Biot-Savart law,
\begin{equation}
    \textbf{B}_\textrm{1D}=\frac{\mu_0I}{4\pi}\int_0^{2\pi}d\tilde \phi\frac{\tilde{\textbf{r}}_c'\times\Delta\textbf{r}_c}{|\Delta\textbf{r}_c|^{3}},\label{100}
\end{equation}
and $\Delta\textbf{r}_c\equiv\textbf{r}_c-\tilde{\textbf{r}}_c$. (Here and throughout the paper, tildes indicate quantities evaluated at the source location, e.g. $\tilde{\mathbf{r}}_c \equiv \mathbf{r}_c(\tilde{\phi})$). The integrand of (\ref{100}) notably contains a nonintegrable singularity at $\tilde{\textbf{r}}_c=\textbf{r}_c$. One may naively attempt to ``cut it out,'' though this does not provide a solution. Consider the known formula for the self-force on a circular conductor, 
\begin{equation}\label{105}
\frac{d\textbf{F}_\circ}{dl}=-\frac{\mu_0I^2}{4\pi R}\left[\ln\left(\frac{8R}{a}\right)-\frac{3}{4}\right]\textbf{e}_2,
\end{equation}
which is obtained from (5.40) of \cite{shafranov1966plasma} and can be derived from the self-inductance (\ref{101}) by the principle of virtual work. (Here and elsewhere, the ``$\circ$'' subscript indicates a formula which pertains to a circular loop). In (\ref{105}), we see that the solution diverges as $a/R\rightarrow 0$,  demonstrating that the singularity meaningfully contributes to the final result: when $a/\mathcal{L}\rightarrow 0$, it is expected for the self-force -- and the magnetic field by extension -- to diverge as well. Therefore, removing the singularity from the domain of integration is not a valid method. This is demonstrated in Figure (\ref{fig:skipping}) for the self-force on a circular loop in which a numerical integration of (\ref{127})-(\ref{100}) where the grid point at $\tilde{\textbf{r}}_c=\textbf{r}_c$ is dropped is compared to other methods introduced in this paper.
When the singular grid point is merely dropped, the resulting force does not converge to any limit as the resolution is increased.

\subsection{Overview of Derivations}

The formulae in our work are determined by rigorously simplifying the three-dimensional integral form of the magnetic vector potential (\ref{9}) under the assumption $a/\mathcal{L}\ll 1$. The key idea in this method is to partition the domain of the integral over $\tilde\phi$ in (\ref{9}) into a ``near region'' in which $|\phi-\tilde\phi|\ll 1$, and a ``far region'' in which $a/|\Delta\textbf{r}|\ll 1$, where $\Delta\textbf{r}\equiv\textbf{r}-\tilde{\textbf{r}}$. This partition occurs at some angle $\phi_0$ such that $a/\mathcal{L}\ll \phi_0\ll 1$. A detailed derivation for the vector potential is given in Appendix \ref{appendix:vecpot}. It is then possible to derive the self-inductance, self magnetic field, and self-force from the vector potential. These derivations are given in Appendices \ref{appendix:induct}-\ref{appendix:force}. 

\subsection{The Self-Inductance}

It is shown in Appendix \ref{appendix:induct} that the self-inductance of thin coils is given by 
\begin{equation}\label{91}
    L=\frac{\mu_0}{4\pi}\int_0^{2\pi}d\tilde\phi\int_0^{2\pi}d\phi\frac{\textbf{r}_c'\cdot\tilde{\textbf{r}}_c'}{\sqrt{|\Delta\textbf{r}_c|^2+\frac{a^2}{\sqrt{e}}}}.
\end{equation}
For a circular conductor of major radius $R$, this becomes
\begin{equation}\label{101}
    L_\circ=\frac{\mu_0R}{2\sqrt{2}}\int_0^{2\pi}d(\Delta\phi)\frac{\cos\Delta\phi}{\sqrt{1-\cos\Delta\phi+\frac{a^2}{2R^2\sqrt{e}}}}.
\end{equation}
We may apply the integral identity
\begin{multline}\label{110}
    \int_0^{2\pi}d\phi\frac{\cos\phi}{\sqrt{1-\cos\phi+\alpha}}=
    \\\frac{4}{\sqrt{2+\alpha}}\left[(1+\alpha)K\left(\frac{2}{2+\alpha}\right)-(2+\alpha)E\left(\frac{2}{2+\alpha}\right)\right],
\end{multline}
where $E(m=k^2)$ and $K(m=k^2)$ are the complete elliptic integrals. Approximating the right-hand side for $\alpha\ll 1$, 
\begin{equation}\label{107}
    \int_0^{2\pi}d\phi\frac{\cos\phi}{\sqrt{1-\cos\phi+\alpha}}\approx\sqrt{2}(-4+5\ln 2-\ln\alpha).
\end{equation}
Applying (\ref{107}) to (\ref{101}), the self-inductance reduces as expected to the known analytic formula (given on page 234 of \cite{jackson1999classical}),
\begin{equation}\label{106}
    L_\circ=\mu_0 R\left[\ln\left(\frac{8R}{a}\right)-\frac{7}{4}\right].
\end{equation}

\begin{figure}[!htb]
        \center{\includegraphics[width=\columnwidth]
        {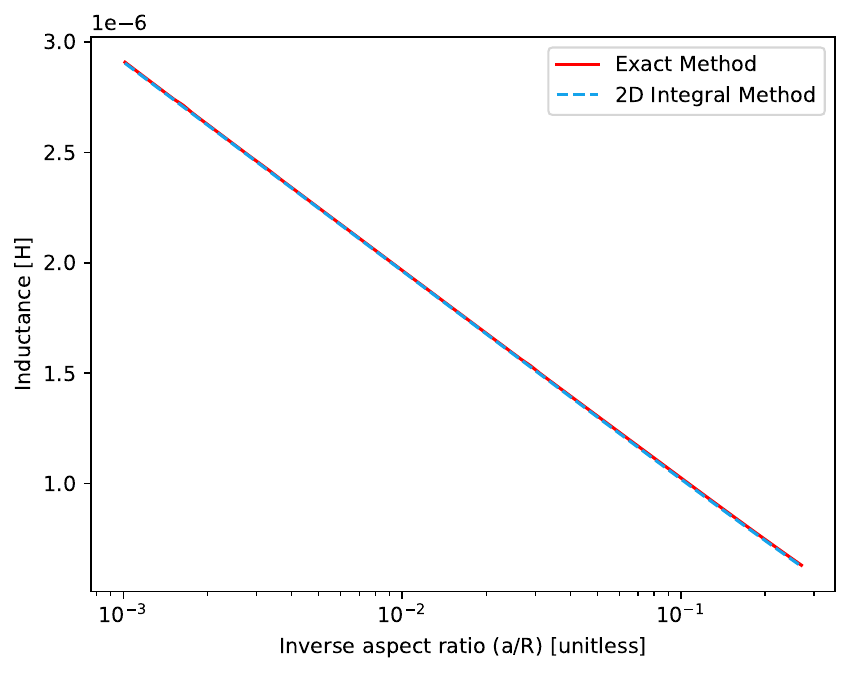}}
        \caption{\label{fig:inductance} The self-inductance of the HSX coil is compared between the exact method (\ref{66}) and the regularized 2D method (\ref{91}) for various values of the inverse aspect ratio $a/R$. These two methods agree extraordinarily well, even for coils where the inverse aspect ratio is slightly below $1$. It is also demonstrated that, in agreement with \ref{122},  the self-inductance is logarithmic in $a/R$. }
        \end{figure}

The regularized self-inductance (\ref{91}) also shows close numerical  agreement with the exact form of the self-inductance (\ref{66}) for the HSX coil in Figure \ref{fig:inductance}. Interestingly, the agreement holds true even for values of $a/R$ that are only slightly below 1, even though (\ref{91})  was was determined under the assumption $a/\mathcal{L}\ll 1$. When the minor radius is varied, the self-inductance depends on $a$ purely through a term proportional to $\ln(l/a)$, where $l$ is the length of the coil: from (\ref{70}) and (\ref{86}), it can be shown that
\begin{multline}\label{122}
    L=\frac{\mu_0l}{2\pi}\ln\left(\frac{l}{a}\right)\\+\frac{\mu_0}{2\pi}\int_0^{2\pi}d\phi|\textbf{r}_c'|\left(\ln\left(\frac{|\textbf{r}_c'|}{l}\right)+\ln\phi_0+\frac{1}{4}\right) +L_{\textrm{reg},f},
\end{multline}
where $L_{\textrm{reg},f}$ is given in (\ref{75}). 
A similar logarithmic dependence on $a$ can be shown to hold for the self-force. 

\subsection{The Self Magnetic Field}

\begin{figure}[!htb]
        \center{\includegraphics[width=\columnwidth]
        {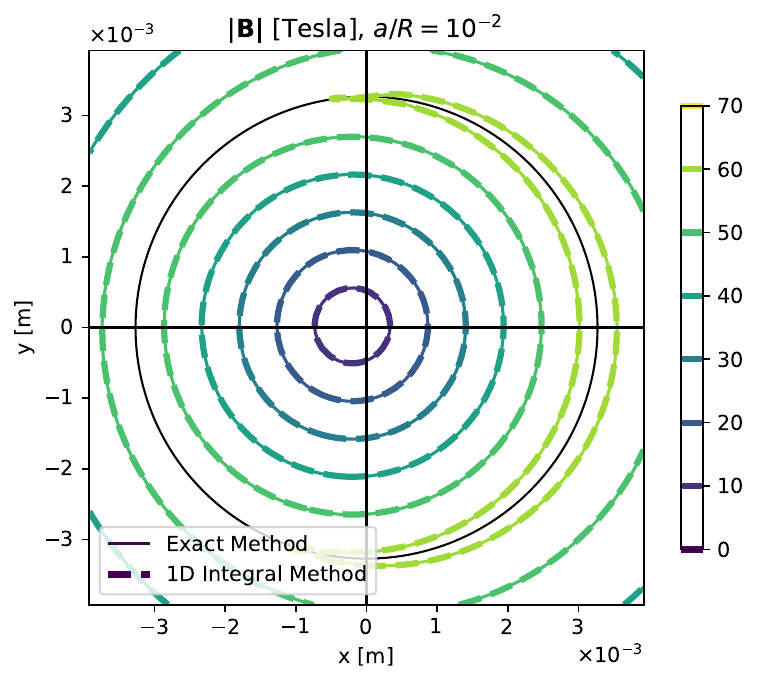}
        \\\includegraphics[width=\columnwidth]
        {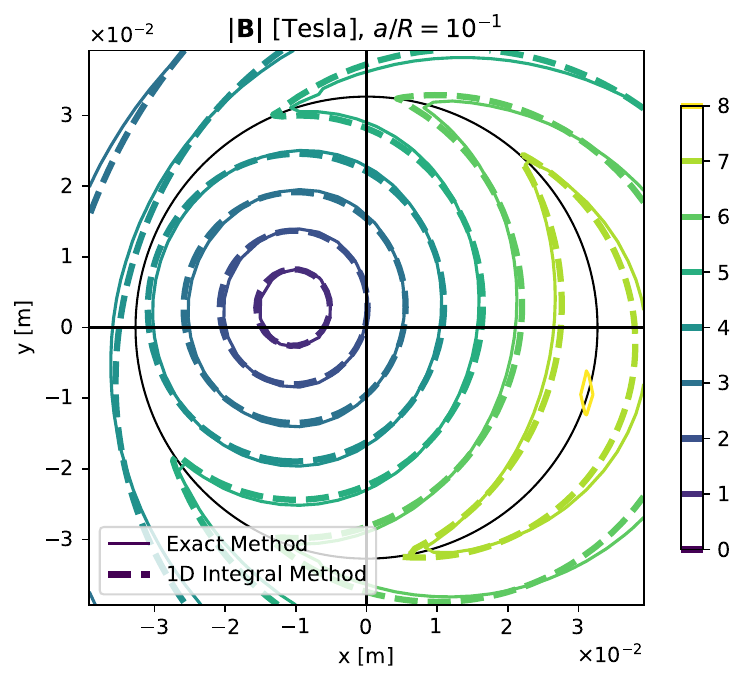}}
        \caption{\label{fig:contours} These contour plots compare the magnitude of the magnetic field in the cross-section of the HSX coil between the exact Biot-Savart law (\ref{109}) and the one-dimensional integral method for $\textbf{B}$ (\ref{102}). 
        The axes are $x=s\cos\theta$ and $y=s\sin\theta$, with the black circle indicating the conductor.
        The plots were evaluated at $\phi=0$ for $I=10^6$ A and for inverse aspect ratios of $10^{-1}$ and $10^{-2}$. These plots demonstrate the convergence of the 1D integral method for $\textbf{B}$ to the exact result with respect to small $a/R$. }
        \end{figure}

It is shown in Appendix \ref{appendix:mag} that the magnetic field for thin coils in the region $s\lesssim a$ is given by
\begin{equation}\label{102}
    \textbf{B}(\textbf{r})=\textbf{B}_\textrm{reg}+\textbf{B}_\textrm{cyl}+\begin{cases}
        \textbf{B}^<, & s\leq a\\
         \textbf{B}^>, & s>a,
    \end{cases}
\end{equation}
where
\begin{equation}\label{43b}
\textbf{B}_\textrm{reg}(\phi)\equiv\frac{\mu_0I}{4\pi}\int_0^{2\pi}d\tilde\phi\frac{\tilde{\textbf{r}}_c'\times\Delta\textbf{r}_c}{\left(|\Delta\textbf{r}_c|^2+\frac{a^2}{\sqrt{e}}\right)^\frac{3}{2}}
\end{equation}
is a regularized Biot-Savart law,
\begin{equation}
    \textbf{B}_\textrm{cyl}(s,\theta,\phi)=\begin{cases}
        \frac{\mu_0Is}{2\pi a^2}(-\sin\theta\textbf{e}_2+\cos\theta\textbf{e}_3), & s\leq a\\
        \frac{\mu_0I}{2\pi s}(-\sin\theta\textbf{e}_2+\cos\theta\textbf{e}_3), & s>a
    \end{cases}
\end{equation}
is the magnetic field of an infinite cylinder of radius $a$ carrying a uniform current $I$, and
\begin{subequations}\label{63}
    \begin{align}
        \begin{split}
            \textbf{B}^<(s,\theta,\phi)&\equiv\frac{\mu_0I\kappa}{8\pi}\bigg[-\frac{s^2\sin(2\theta)\textbf{e}_2}{2a^2}
            \\&+\left(\frac{3}{2}+\frac{s^2}{a^2}\left(\frac{\cos(2\theta)}{2}-1\right)\right)\textbf{e}_3\bigg]
        \end{split}
        \\ \begin{split}
            \textbf{B}^>(s,\theta,\phi)&\equiv\frac{\mu_0I\kappa}{8\pi}\left[\left(\frac{a^2\sin (2\theta)}{2s^2}-\sin (2\theta)\right)\textbf{e}_2\right.
    \\& \left.+\left(\frac{1}{2}-2\ln\left(\frac{s}{a}\right) -\frac{a^2\cos (2\theta)}{2s^2}+\cos (2\theta)\right)\textbf{e}_3\right].
        \end{split}
    \end{align}
\end{subequations}
Figure \ref{fig:contours} demonstrates the convergence of the one-dimensional magnetic field formula (\ref{102}) to the exact Biot-Savart law,
\begin{equation}\label{109}
    \textbf{B}(\textbf{r})=\frac{\mu_0I}{4\pi^2 a^2}\int_0^{2\pi}\hspace{-0.5em}d\tilde\phi\int_0^{2\pi}\hspace{-0.5em}d\tilde\theta\int_0^a d\tilde s\,\tilde s (1-\tilde\kappa \tilde s\cos\tilde\theta)\frac{\tilde{\textbf{r}}_c'\times\Delta\textbf{r}}{|\Delta\textbf{r}|^{3}},
\end{equation}
across contours of $\textbf{B}$ on the HSX coil at different aspect ratios. As expected, the two formulae show good agreement, with closer agreement at higher aspect ratios. 

\subsection{The Self-Force}
\begin{figure}[!htb]
        \includegraphics[width=\columnwidth]
        {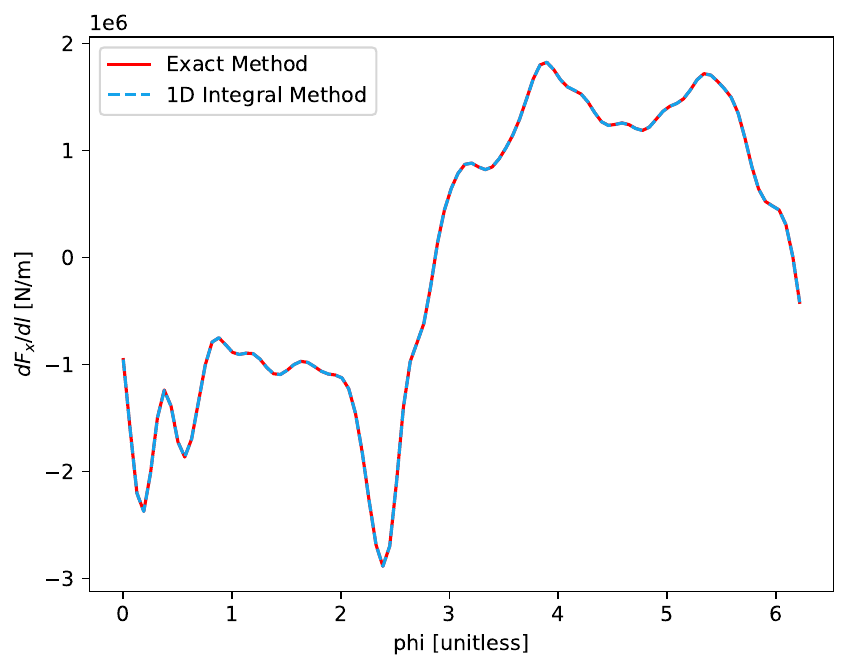}
        \caption{\label{fig:self-force}The self-force in the $\hat{\textbf{x}}$-direction for the HSX coil displays excellent agreement between the exact method (\ref{123}) and the one-dimensional integral formula (\ref{103}). The current was taken to be $10^6$ A and the inverse aspect ratio $a/R$ was taken to be $10^{-2}$. }
        \end{figure}

It is shown in Appendix \ref{appendix:force} that the self-force for thin coils is given by
\begin{equation}\label{103}
    \frac{d\textbf{F}}{dl}=I\textbf{e}_1\times\textbf{B}_\textrm{reg}.
\end{equation}
In an identical fashion to the self-inductance of a circular coil (\ref{101}-\ref{106}), it can be shown that a circular coil's self-force reduces to the analytically-derived expression (\ref{105}). The one-dimensional integral formula (\ref{103}) and the exact form of the self-force (\ref{123}) also show strong numerical agreement for the HSX coil, as shown in Figure \ref{fig:self-force}. Just as with the self-inductance, this agreement holds true even at low aspect ratio, despite the assumption $a/\mathcal{L}\ll 1$ in the analysis. 
\subsection{The Maximum Magnetic Field Strength}

An additional quantity of interest is $\max_{s,\theta}|\textbf{B}|$ within the coil as it determines the critical current and temperature for a superconductor. Although the current density in superconductors is nonuniform, the value of $\max_{s,\theta}|\textbf{B}|$ computed for a uniform current density may be a good enough approximation (particularly in multi-coil systems) to use for optimization.

Unfortunately, even with the uniform current density approximation, (\ref{102}) is too complicated for the maximum over $(s,\theta)$ to be determined analytically. However, we may arrive at a reasonable approximation by noting that $\textbf{B}^<$ is formally smaller in magnitude than $\textbf{B}_\textrm{cyl}$ and generally smaller than $\textbf{B}_\textrm{reg}$, a fact that can be demonstrated for a circular coil. Therefore, it is reasonable to approximate the location of $\max_{s,\theta}|\textbf{B}|$ as the point that instead maximizes the quantity $\textbf{B}^{(0)}\equiv\textbf{B}_\textrm{reg}+\textbf{B}_\textrm{cyl}$. This point is found to be given by $s_\textrm{max}=a$ and $\theta_\textrm{max}=-\tan^{-1}(B_\textrm{reg,2}/B_\textrm{reg,3})$, where $B_\textrm{reg,j}\equiv\textbf{B}_\textrm{reg}\cdot\textbf{e}_j$. Substituting this solution into (\ref{102}) and keeping terms up to zeroth order in $a/\mathcal{L}$,
\begin{equation}\label{114}
    \max_{s,\theta}|\textbf{B}|^2\approx B_{\textrm{reg,1}}^2+B^2_\parallel\left(1+\frac{1}{B_\parallel}\frac{\mu_0I}{2\pi a}\right)^2,
\end{equation}
where $B_\parallel^2\equiv B^2_{\textrm{reg,2}}+B^2_{\textrm{reg,3}}$. 

\begin{figure}[!htb]
        \center{\includegraphics[width=\columnwidth]
        {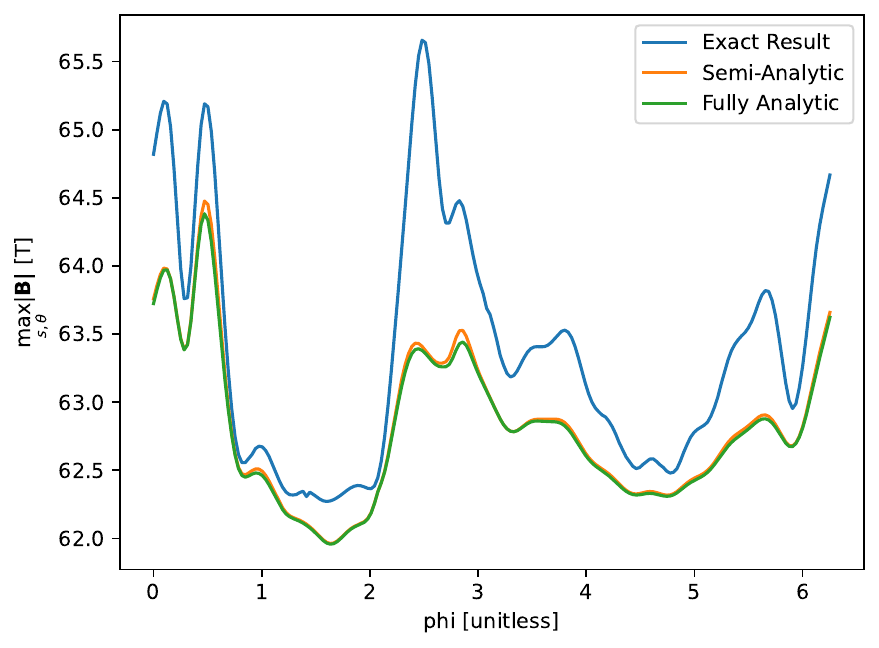}}
        \caption{\label{fig:modb} This plot compares $\max_{s,\theta}|\textbf{B}|$ along the length of the HSX coil between an exact result obtained from numerically maximizing (\ref{109}), a semi-analytic result obtained from numerically maximizing (\ref{102}), and an analytic result (\ref{114}). The coil was chosen to carry a current of $10^6$ A and have an inverse aspect ratio, $a/R$, of $10^{-1}$.} 
        \end{figure}

The accuracy of (\ref{114}) is demonstrated in Figure \ref{fig:modb} by comparing it to numerical maximizations of the exact Biot-Savart law (\ref{109}) and one-dimensional magnetic field calculation (\ref{102}) for the HSX coil. The analytic maximization (\ref{114}) agrees closely with the numerical maximization of (\ref{102}), which indicates that the decision to find the maximum of $|\textbf{B}^{(0)}|$ instead of $|\textbf{B}|$ was reasonable. The agreement between the exact Biot Savart law (\ref{109}) and the one-dimensional methods is weaker, though it may be sufficiently close for (\ref{114}) to be useful for estimates and within optimization.

\section{Efficient Quadrature}\label{sec:quad}
\begin{figure}
        \includegraphics[width=\columnwidth]
        {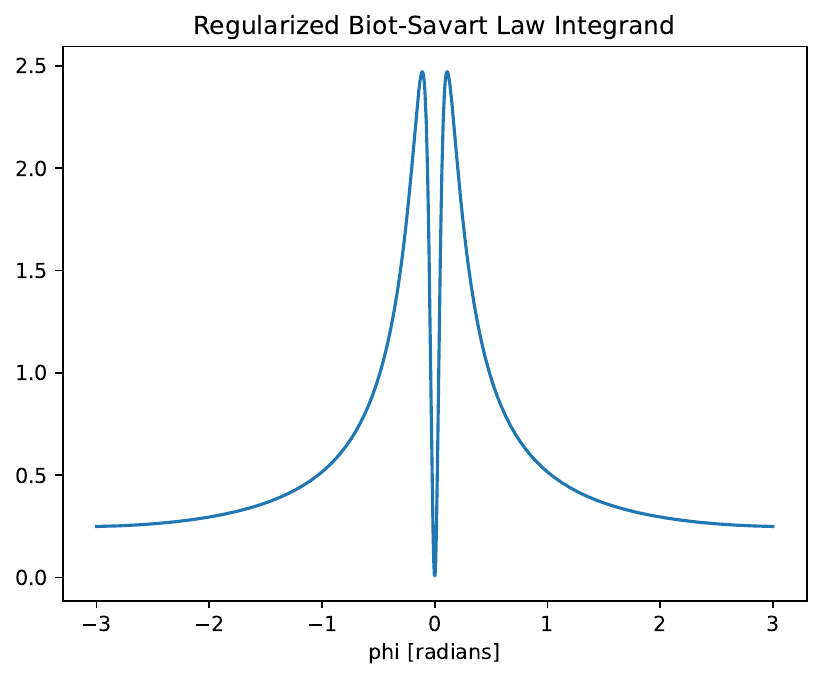}
        \caption{\label{fig:sharp}The integrand of the regularized Biot-Savart law (\ref{43b}) is shown for a circle with $a/{R}=1/10$, where $R$ is the major radius. This demonstrates the fine-scale structure of the integrand near $\tilde{\textbf{r}}_c\approx \textbf{r}_c$. }\end{figure}

A key advantage of the simplified formulae for the self-inductance (\ref{91}) and self-force (\ref{103}) is that neither contain a singularity at $\tilde{\textbf{r}}_c=\textbf{r}_c$. Even so, quadrature is still challenging as the integrands have fine-scale structure near $\tilde{\textbf{r}}_c\approx \textbf{r}_c$ for $a/\mathcal{L} \ll 1$. Figure \ref{fig:sharp} demonstrates the fine-scale structure of the integrand of the regularized Biot-Savart law (\ref{43b}).

One method to circumvent this issue relies upon subtracting and adding back some analytically integrable function, $\boldsymbol{\Lambda}(\phi,\tilde\phi)$, that fits the integrand well near the reference point. For the regularized Biot-Savart law, we choose $\boldsymbol{\Lambda}$ based on the integrand of  (\ref{43b}) as follows. First, we expand the numerator and denominator about $\tilde\phi=\phi$ using the result of (\ref{19}) and keep terms to second order in $\Delta\phi$. Comparing this expression to (\ref{3a}), we find
\begin{equation}
    \boldsymbol{\Lambda}=\frac{\kappa(\Delta\phi)^2\textbf{e}_3}{2\left((\Delta\phi)^2+\frac{a^2}{|\textbf{r}_c'|^2\sqrt{e}}\right)^\frac{3}{2}}.
\end{equation}
In order to recapture the $2\pi$-periodic behavior of the integrand of (\ref{43b}), we let $(\Delta\phi)^2\rightarrow 2(1-\cos\Delta\phi)$, giving
\begin{equation}\label{111}
    \boldsymbol{\Lambda} = \frac{\kappa(1-\cos\Delta\phi)\textbf{e}_3}{2^{3/2}\left((1-\cos\Delta\phi)+\frac{a^2}{2|\textbf{r}_c'|^2\sqrt{e}}\right)^\frac{3}{2}}.
\end{equation}
Comparing to the formulae for self-force of a circular coil (\ref{105}, \ref{103}), we see
\begin{equation}\label{113}
    \frac{\mu_0I}{4\pi}\int_0^{2\pi}d\tilde\phi\boldsymbol{\Lambda}=\frac{\mu_0I\kappa}{4\pi}\left(\ln\left(\frac{8|\textbf{r}_c'|}{a}\right)-\frac{3}{4}\right)\textbf{e}_3.
\end{equation}
The advantage of (\ref{111}) is that it allows us to subtract $\boldsymbol{\Lambda}$ from the integrand of (\ref{43b}) and add it back to obtain a \textit{modified} regularized Biot-Savart law,

\begin{multline}\label{108}
    \textbf{B}_\textrm{reg}=\frac{\mu_0I}{4\pi}\hspace{-0.25em}\int_0^{2\pi}\hspace{-0.5em}d\tilde\phi\left(\frac{\tilde{\textbf{r}}_c'\times\Delta\textbf{r}_c}{\left(|\Delta\textbf{r}_c|^2+\frac{a^2}{\sqrt{e}}\right)^{\frac{3}{2}}}-\boldsymbol{\Lambda}\right)
    \\+\frac{\mu_0I\kappa}{4\pi}\left(\ln\left(\frac{8|\textbf{r}_c'|}{a}\right)-\frac{3}{4}\right)\textbf{e}_3,
\end{multline}
where we applied (\ref{113}). Although algebraically identical to (\ref{43b}), (\ref{108}) is easier to evaluate as the new integral has a significantly smoother integrand at $\tilde{\textbf{r}}_c\approx\textbf{r}_c$. 

\begin{figure}[!htb]
        \includegraphics[width=\columnwidth]
        {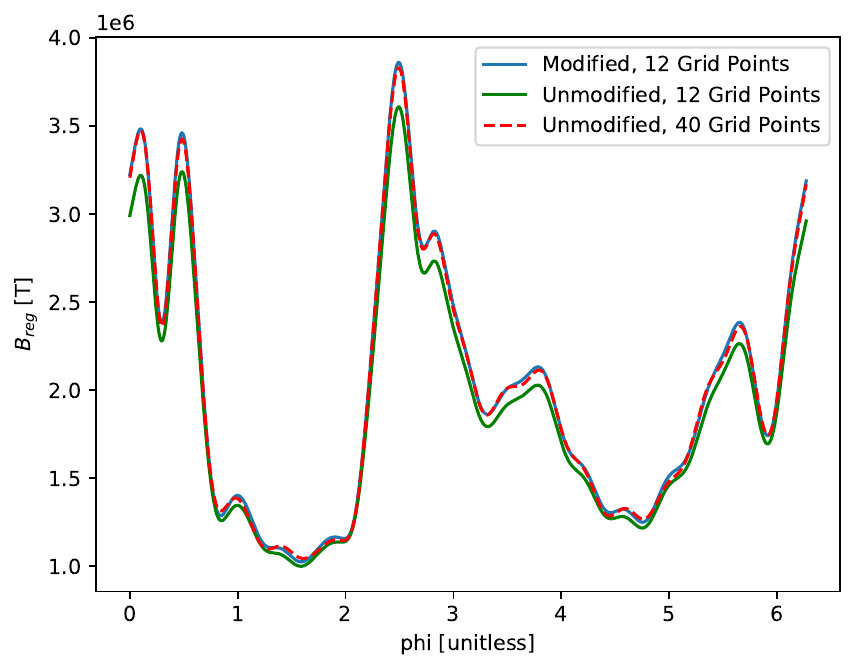}
        \caption{\label{fig:regbiotsavart} This plot compares evaluations of the modified (\ref{108}) and unmodified (\ref{43b}) regularized Biot-Savart laws across the length of the HSX coil for various choices of quadrature grid resolution in order to demonstrate relative rate of convergence. The modified method performs significantly better than the unmodified method, a 12-point grid for the modified method yielded comparable results to a 40-point grid for the unmodified method. The current is taken to be $I=10^6$ A and the inverse aspect ratio $a/R$ is $10^{-2}$. }
        \end{figure}

The efficiency of this modified method in comparison to the unmodified regularized Biot-Savart law (\ref{43b}) is demonstrated in Figure \ref{fig:regbiotsavart}, where the results of both methods are shown for several quadrature grid resolutions. For each value of $\phi$ in the figure, $\textbf{B}_\textrm{reg}$ is computed using Gauss-Legendre quadrature on the interval $\tilde{\phi}\in [\phi, \phi+2\pi]$. Compared to highly resolved calculations, the modified method achieves accuracy within $1\%$ for a grid of only 12 points, while the unmodified method requires 40 points to do the same. For uniformly spaced quadrature points, similar results are obtained with a slightly larger number of points.

\begin{figure}[!htb]
        \includegraphics[width=\columnwidth]
        {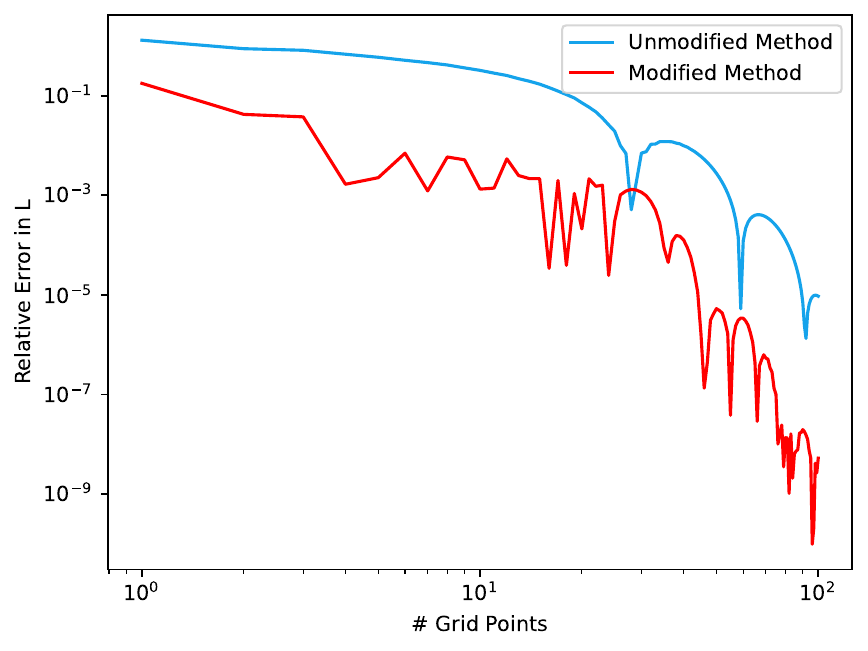}s
        \caption{\label{fig:L_conv}This plot compares evaluations of the modified (\ref{124}) and unmodified (\ref{91}) 2D integral forms of the self-inductance for the HSX coil at various choices of quadrature grid resolution in order to demonstrate relative rate of convergence. The coil was taken to have an inverse aspect ratio $a/R$ of $10^{-2}$.}
        \end{figure}

Using the same methodology, a modified expression can be obtained for the self-inductance,
\begin{multline}\label{124}
    L = \frac{\mu_{0}}{4\pi}\int_0^{2\pi} d\phi\left|\mathbf{r}_{c}'\right|\left(2\ln\left(\frac{8\left|\mathbf{r}_{c}'\right|}{a}\right)+\frac{1}{2}\right)\\
  +\frac{\mu_{0}}{4\pi}\int_0^{2\pi} d\phi\int_0^{2\pi} d\tilde{\phi}\left(\frac{\mathbf{r}_{c}'\cdot\tilde{\mathbf{r}}_{c}'}{\sqrt{\left|\Delta\textbf{r}_c\right|^{2}+\frac{a^2}{\sqrt{e}}}}
 \right. \\ \left.
 -\frac{\left|\mathbf{r}_{c}'\right|^{2}}{\sqrt{2(1-\cos\Delta\phi)\left|\mathbf{r}_{c}'\right|^{2}+\frac{a^2}{\sqrt{e}}}}\right).
\end{multline}
Figure \ref{fig:L_conv} demonstrates the efficiency of this method in comparison to the unmodified 2D integral form of the self-inductance (\ref{91}) by comparing their predictions for the self-inductance of the HSX coil at various choices of quadrature grid resolution. 

\section{Conclusion}

The self-inductance, self magnetic field, and self-force of electromagnetic coils are important properties yet they are difficult to evaluate numerically due to the high dimensionality of the integrals and the presence of a singularity in the integrands. In this paper, we rigorously derived novel integral formulae for these quantities using an expansion in large coil aspect ratio.
The main results are summarized in Section \ref{sec:simplifiedform},
with (\ref{91}) for the self-inductance, (\ref{102})-(\ref{63}) for the field, (\ref{103}) for the self-force, and (\ref{114}) for the maximum field strength in the conductor.
These formulae are described by integrals of significantly reduced dimensionality  that no longer contain a singularity. We demonstrated that these formulae show remarkable agreement to the ``exact'' finite-thickness integrals even at low aspect ratios, both analytically for a torus and numerically for the HSX coil. In Section \ref{sec:quad}, we subsequently derived expressions to compute the self-inductance (\ref{124}) and regularized Biot-Savart law (\ref{108}) with even greater efficiency, and demonstrated that the self-force can be accurately computed with a grid of as few as 12 points. 

A key direction of future research would be to incorporate measures of stress and critical current, both of which may be computed from our simplified integral formulae, into coil shape optimization. For application to fusion energy, the design of stellarators in particular may benefit from this work as there is significant flexibility in coil shape. Prior to this work, it would have been infeasible to optimize for stress or critical current as the calculations would have been too costly. The modified forms of the self-inductance and regularized Biot-Savart law  now allow for optimization as they can be evaluated rapidly. An additional direction for research would be to generalize these results for electromagnetic coils with different geometries. 
For example, a similar methodology can be applied to coils in which the cross-section is rectangular, as will be discussed in a separate paper  \cite{PaperII}. 

\appendices

\section{Reduced Integral Formula for the Magnetic Vector Potential\label{appendix:vecpot}}
The magnetic vector potential is given by a three-dimensional integral,
\begin{equation}\label{126}
    \textbf{A}=\frac{\mu_0}{4\pi}\int d^3\tilde{\textbf{r}}\frac{\tilde{\textbf{J}}}{|\Delta\textbf{r}|}.
\end{equation}
We consider a closed conductor with a circular cross-section of minor radius $a$, carrying a uniformly distributed current $I$ in the $\textbf{e}_1$ direction, (\ref{130}). In our coordinates, the magnetic vector potential for this coil is given by,
\begin{equation}\label{9}
    \textbf{A}(\textbf{r})=\frac{\mu_0I}{4\pi^2a^2}\int_0^{2\pi}\hspace{-0.5em}d\tilde{\phi}\int_0^{2\pi}\hspace{-0.5em}d\tilde{\theta}\int_0^a d\tilde{s} \, \tilde{s}(1-\tilde\kappa \tilde s\cos\tilde\theta)\frac{\tilde{\textbf{r}}_c'}{|\Delta \textbf{r}|}, 
\end{equation}
where $\Delta\textbf{r}\equiv \textbf{r}-\tilde{\textbf{r}}$. In this Appendix, we show that when $s\lesssim a$ and the minor radius is significantly smaller than the length scales $\mathcal{L}$ associated with $\textbf{r}_c$ (which includes $|\textbf{r}_c'|$, $\kappa^{-1}$, and $\tau^{-1}$), there exists a reduced form of the magnetic vector potential,
\begin{multline}\label{10}
    \textbf{A}=\frac{\mu_0I}{4\pi}\hspace{-2em}\int\limits_{\hspace{1em}\phi+\phi_0}^{\hspace{1em}2\pi+\phi-\phi_0}\hspace{-1.5em}d\phi'\frac{\textbf{r}_c'}{|\Delta \textbf{r}_c|}\left[1-\frac{\Delta \textbf{r}_c\cdot s(\cos\theta\textbf{e}_2+\sin\theta\textbf{e}_3)}{|\Delta \textbf{r}_c|^2}\right]
    \\+\frac{\mu_0I\textbf{e}_1}{2\pi}\left(1+\frac{\kappa s\cos\theta}{2}\right)\left(\ln\left(\frac{|\textbf{r}_c'|}{a}\right)+\ln(2\phi_0)\right)
    \\+\frac{\mu_0I\textbf{e}_1}{4\pi }\begin{cases}
        -\frac{\kappa a^2\cos\theta}{4s}-\kappa s\cos\theta, & s>a\\
        (1-\kappa s\cos\theta)-\frac{s^2}{a^2}\left(1+\frac{\kappa s\cos\theta}{4}\right), & s\leq a
    \end{cases},
\end{multline}
where $\phi_0$ is defined ahead. This formula will be important in Appendices \ref{appendix:induct}, \ref{appendix:mag}, and \ref{appendix:force} for finding simplified integral forms of the self-inductance, magnetic field, and self-force, respectively. 

To reduce (\ref{9}) for $a/\mathcal{L}\ll 1$, different simplifying assumptions can be applied in different regions of the domain. In the ``far region'' where $|\phi-\tilde\phi|\sim 1$ and $a\ll|\Delta \textbf{r}|$, the three-dimensional structure of the conductor is negligible and the coil can be approximated by a one-dimensional filament. By contrast, in the ``near region'' where $|\Delta \textbf{r}|\sim a$ and $|\phi-\tilde\phi|\ll 1$, the three-dimensional structure is not negligible and must be considered. In order to take both domains into account, we introduce an intermediate angle $\phi_0$ such that $a/\mathcal{L}\ll \phi_0\ll 1$. Then, we split the integral over $\tilde\phi$ in (\ref{9}) in two: the near region  $[\phi-\phi_0, \phi+\phi_0]$ (in which $|\phi-\tilde\phi|\ll 1$) and the far region  $[\phi+\phi_0, 2\pi+\phi-\phi_0]$ (in which $a/|\Delta \textbf{r}|\ll 1$).
%The final result should be independent of $\phi_0$. 
In other words, we write 
\begin{equation}\label{11}
    \textbf{A}=\textbf{A}_n+\textbf{A}_f,
\end{equation}
where
\begin{subequations}\label{12}
    \begin{gather}
        \textbf{A}_n=\frac{\mu_0I}{4\pi^2a^2}\hspace{-0.5em}\int\limits_{\phi-\phi_0}^{\phi+\phi_0}\hspace{-0.5em}d\tilde{\phi}\int_0^{2\pi}\hspace{-0.5em}d\tilde{\theta}\int_0^a d\tilde{s} \tilde{s}(1-\tilde\kappa \tilde s\cos\tilde\theta)\frac{\tilde{\textbf{r}}_c'}{|\Delta \textbf{r}|}\label{12a}
        \\\textbf{A}_f=\frac{\mu_0I}{4\pi^2a^2}\hspace{-2em}\int\limits_{\hspace{1em}\phi+\phi_0}^{\hspace{1em}2\pi+\phi-\phi_0}\hspace{-1.5em}d\tilde{\phi}\int_0^{2\pi}\hspace{-0.5em}d\tilde{\theta}\int_0^a d\tilde{s} \tilde{s}(1-\tilde\kappa \tilde s\cos\tilde\theta)\frac{\tilde{\textbf{r}}_c'}{|\Delta \textbf{r}|}.\label{12b}
    \end{gather}
\end{subequations}
A similar idea of identifying an intermediate scale for the self-inductance was used on page 123 of \cite{landau2013electrodynamics} and in \cite{dengler2012self}. 
We proceed to simplify these integrals in each region.

\subsection{Far Region}
From the position vector (\ref{4}),
\begin{equation}\label{121}
\Delta\textbf{r}=\Delta\textbf{r}_c+s\cos\theta\textbf{e}_2+s\sin\theta\textbf{e}_3
    -\tilde s\cos\tilde\theta\tilde{\textbf{e}}_2-\tilde s\sin\tilde\theta\tilde{\textbf{e}}_3,
\end{equation}
where $\Delta\textbf{r}_c\equiv\textbf{r}_c - \tilde{\textbf{r}}_c$. Performing a Taylor expansion in $a/\mathcal{L}$ and keeping terms up to first order,
\begin{multline}\label{14}
    \frac{1}{|\Delta \textbf{r}|}=\frac{1}{|\Delta \textbf{r}_c|}\left[1-\frac{\Delta \textbf{r}_c}{|\Delta \textbf{r}_c|^2}\cdot(s\cos\theta\textbf{e}_2+s\sin\theta\textbf{e}_3
    \right.\\\left.-\tilde s\cos\tilde\theta\tilde{\textbf{e}}_2-\tilde s\sin\tilde\theta\tilde{\textbf{e}}_3)\right].
\end{multline}
Next, we substitute (\ref{14}) into (\ref{12b}), perform the integral over $\tilde\theta$, and keep terms up to first order in $a/\mathcal{L}$. Performing the next integral over $\tilde s$, 
\begin{equation}\label{16}
    \textbf{A}_f=\frac{\mu_0I}{4\pi}\hspace{-2em}\int\limits_{\hspace{1em}\phi+\phi_0}^{\hspace{1em}2\pi+\phi-\phi_0}\hspace{-1.5em}d\tilde{\phi}\frac{\tilde{\textbf{r}}_c'}{|\Delta \textbf{r}_c|}\bigg[1-
   \frac{\Delta \textbf{r}_c}{|\Delta \textbf{r}_c|^2}\cdot(s\cos\theta\textbf{e}_2+s\sin\theta\textbf{e}_3)\bigg].
\end{equation}
Comparing this expression to (\ref{14}), we note that (\ref{16}) may more simply be written as
\begin{equation}\label{162}
    \textbf{A}_f=\frac{\mu_0I}{4\pi}\hspace{-2em}\int\limits_{\hspace{1em}\phi+\phi_0}^{\hspace{1em}2\pi+\phi-\phi_0}\hspace{-1.5em}d\tilde{\phi}\frac{\tilde{\textbf{r}}_c'}{|\textbf{r}-\tilde{\textbf{r}}_c|}.
\end{equation}

\subsection{Near Region\label{appendix:vecpotnear}}
In the near region, we use the assumption that $|\Delta\phi|\ll 1$, where $\Delta\phi\equiv \phi-\tilde\phi$. To begin, we expand $|\Delta \textbf{r}|$ for small $\Delta\phi$. First, we perform a Taylor expansion of $\textbf{r}_c$ to second order in $\Delta\phi$,
\begin{equation}\label{19}
    \Delta\textbf{r}_c=\left(|\tilde{\textbf{r}}_c'|\Delta\phi+\frac{1}{2}\frac{d|\tilde{\textbf{r}}_c'|}{d\tilde\phi}(\Delta\phi)^2\right)\tilde{\textbf{e}}_1+\frac{\tilde\kappa|\tilde{\textbf{r}}_c'|^2(\Delta\phi)^2}{2}\tilde{\textbf{e}}_2,
\end{equation}
where we determined $\tilde{\textbf{r}}_c''$ by evaluating the $\textbf{e}_1'$ term of (\ref{2}) with (\ref{1}). Substituting (\ref{19}) into (\ref{121}), performing Taylor expansions of $\textbf{e}_2$ and $\textbf{e}_3$, and keeping terms up to second order in $\Delta\phi$, 
\begin{multline}\label{20}
    \Delta\textbf{r}=\left(|\tilde{\textbf{r}}_c'|\Delta\phi+\frac{1}{2}\frac{d|\tilde{\textbf{r}}_c'|}{d\tilde\phi}(\Delta\phi)^2\right)\tilde{\textbf{e}}_1
    \\+\left(\frac{\tilde\kappa|\tilde{\textbf{r}}_c'|^2(\Delta\phi)^2}{2}+s\cos\theta-\tilde s\cos\tilde\theta\right)\tilde{\textbf{e}}_2
    +(s\sin\theta-\tilde s\sin\tilde\theta)\tilde{\textbf{e}}_3
    \\+s\cos\theta\Delta\phi\left(\tilde{\textbf{e}}_2'+\frac{\tilde{\textbf{e}}_2''\Delta\phi}{2}\right)
    +s\sin\theta\Delta\phi\left(\tilde{\textbf{e}}_3'+\frac{\tilde{\textbf{e}}_3''\Delta\phi}{2}\right),
\end{multline}
where the derivatives of the basis vectors can be found from the Frenet-Serret formulas (\ref{2}). We assume all length scales $\mathcal{L}$ of $\textbf{r}_c$ in (\ref{2}) are comparable, so $\textbf{e}_i \sim d\textbf{e}_i/d\phi \sim d^2\textbf{e}_i/d\phi^2$, where $\textbf{e}_i$ is any one of the Frenet-Serret basis vectors. As a result, we can neglect the terms in (\ref{20}) containing second derivatives of the basis vectors; these terms are second order in $\Delta\phi$ and are additionally one order higher in the small quantity $a/\mathcal{L}$ than the remaining $(\Delta\phi)^2$ terms. Neglecting these second derivative terms and using (\ref{2}) to evaluate the first derivative terms, we find
\begin{multline}\label{21}
    \Delta\textbf{r}=\left((1-\tilde\kappa s\cos\theta)|\tilde{\textbf{r}}_c'|\Delta\phi+\frac{1}{2}\frac{d|\tilde{\textbf{r}}_c'|}{d\tilde\phi}(\Delta\phi)^2\right)\tilde{\textbf{e}}_1
    \\+\left(-\tilde s\cos\tilde\theta+s\cos\theta-\tilde\tau s\sin\theta|\tilde{\textbf{r}}_c'|\Delta\phi+\frac{\tilde\kappa|\tilde{\textbf{r}}_c'|^2(\Delta\phi)^2}{2}\right)\tilde{\textbf{e}}_2
    \\+\left(-\tilde s\sin\tilde\theta+s\sin\theta+\tilde\tau s\cos\theta|\tilde{\textbf{r}}_c'|\Delta\phi\right)\tilde{\textbf{e}}_3.
\end{multline}
Squaring this result; expanding $|\tilde{\textbf{r}}_c'|$, $\tilde\kappa$, and $\tilde \tau$ about $\tilde\phi=\phi$; and keeping all terms up to second order in $\Delta\phi$ and first order in $a/\mathcal{L}$, 
\begin{multline}\label{23}
    |\Delta\textbf{r}|^2=(s^2+\tilde s^2-2s\tilde s\cos\Delta\theta)+2\tau s\tilde s|\textbf{r}_c'|\sin(\Delta\theta)\Delta\phi
    \\+(1-\kappa s\cos\theta-\kappa\tilde s\cos\tilde\theta)|\textbf{r}_c'|^2(\Delta\phi)^2,
\end{multline}
where $\Delta\theta\equiv\theta-\tilde\theta$. It is helpful to abbreviate this expression as
\begin{equation}\label{24}
    |\Delta\textbf{r}|^2=\alpha+\beta\Delta\phi+\gamma(\Delta\phi)^2,
\end{equation}
where
\begin{subequations}\label{25}
    \begin{gather}
       \alpha=s^2+\tilde s^2-2s\tilde s\cos\Delta\theta\\
       \beta=2\tau s\tilde s|\textbf{r}_c'|\sin\Delta\theta\\
       \gamma=(1-\kappa s\cos\theta-\kappa\tilde s\cos\tilde\theta)|\textbf{r}_c'|^2.
    \end{gather}
\end{subequations}
Note that $\alpha,\beta\sim a^2$ while $\gamma\sim \mathcal{L}^2$. 

By expanding $\tilde{\textbf{r}}_c'$ and $\tilde\kappa$ about $\tilde\phi=\phi$ and applying (\ref{23}), (\ref{12a}) can be expressed in terms of integrals of the form
\begin{equation}\label{262}
    \int_{-\phi_0}^{\phi_0}d(\Delta\phi)\frac{(\Delta\phi)^m}{|\Delta\textbf{r}|},
\end{equation}
where $m$ is an integer $\geq 0$. Let us seek to simplify the first few cases of (\ref{262}). For $m=0$, \cite{gradshteyn2014table}
\begin{equation}\label{27}
    \int \frac{d(\Delta\phi)}{|\Delta\textbf{r}|}=\frac{\ln\left(2\sqrt{\gamma}\sqrt{\alpha+\beta \Delta\phi+\gamma (\Delta\phi)^2}+2\gamma \Delta\phi+\beta\right)}{\sqrt{\gamma}}.
\end{equation}
    As $\alpha$ and $\beta\sim a^2$ while $\gamma\sim \mathcal{L}^2$ in (\ref{25}), the $m=0$ case simplifies under the assumption $a/\mathcal{L}\ll\phi_0\ll 1$ to
\begin{equation}\label{28}
    \int_{-\phi_0}^{\phi_0}\frac{d(\Delta\phi)}{|\Delta\textbf{r}|}\approx \frac{1}{\sqrt{\gamma}}\ln\left(\frac{4\gamma\phi_0^2}{\alpha}\right). 
\end{equation}
Substituting in (\ref{25}) and performing a Taylor expansion on $\sqrt{\gamma}$,
\begin{multline}\label{30}
     \int_{-\phi_0}^{\phi_0}\frac{d(\Delta\phi)}{|\Delta\textbf{r}|}=\frac{1}{|\textbf{r}_c'|}\left(1+\frac{\kappa s\cos\theta+\kappa\tilde s\cos\tilde\theta}{2}\right)
     \\\times\ln\left(\frac{4\phi_0^2(1-\kappa s\cos\theta-\kappa\tilde s\cos\tilde\theta)|\textbf{r}_c'|^2}{s^2+\tilde s^2-2s\tilde s\cos\Delta\theta}\right).
\end{multline}
For the $m=1$ and $m=2$ cases,  \cite{gradshteyn2014table} gives
\begin{subequations}
    \begin{gather}
        \int d(\Delta\phi)\frac{\Delta\phi}{|\Delta\textbf{r}|}=\frac{|\Delta\textbf{r}|}{\gamma}-\frac{\beta}{2\gamma}\int\frac{d(\Delta\phi)}{|\Delta\textbf{r}|}\\
        \begin{split}\int d(\Delta\phi)\frac{(\Delta\phi)^2}{|\Delta\textbf{r}|}=\left(\frac{\Delta\phi}{2\gamma}-\frac{3\beta}{4\gamma^2}\right)|\Delta\textbf{r}|\\+\left(\frac{3\beta^2}{8\gamma^2}-\frac{\alpha}{2\gamma}\right)\int\frac{d(\Delta\phi)}{|\Delta\textbf{r}|},\end{split}
    \end{gather}
\end{subequations}
which, under the assumptions that $a/\mathcal{L}\ll\phi_0\ll 1$, yield
\begin{subequations}\label{32}
    \begin{gather}
        \int_{-\phi_0}^{\phi_0} d(\Delta\phi)\frac{\Delta\phi}{|\Delta\textbf{r}|}\approx\frac{\beta}{\gamma^\frac{3}{2}}\left(1-\frac{1}{2}\ln\left(\frac{4\gamma\phi_0^2}{\alpha}\right)\right)
        \\ \int_{-\phi_0}^{\phi_0} d(\Delta\phi)\frac{(\Delta\phi)^2}{|\Delta\textbf{r}|}\approx\frac{\phi_0^2}{\sqrt{\gamma}}-\frac{\alpha}{2\gamma^\frac{3}{2}}\ln\left(\frac{4\gamma\phi_0^2}{\alpha}\right).
    \end{gather}
\end{subequations}
Notably, the $m=1$ and $m=2$ integrals (\ref{32}) are smaller than (\ref{28}) by factors of $(a/\mathcal{L})^2$ and $\phi_0^2$. We therefore use this observation to argue that only the $m=0$ integral (\ref{28}) contributes meaningfully to an evaluation of (\ref{12a}). As a result, in (\ref{12a}) we can expand $\tilde{\textbf{r}}_c'$ and $\tilde\kappa$ about $\tilde\phi=\phi$ and only keep the zeroth order term of $\Delta\phi$, 
\begin{equation}\label{33}
    \textbf{A}_n=\frac{\mu_0I|\textbf{r}_c'|\textbf{e}_1}{4\pi^2a^2}\int_0^{2\pi}d\tilde\theta\int_0^a d\tilde s\tilde s(1-\kappa\tilde s\cos\tilde\theta)\int_{-\phi_0}^{\phi_0}\frac{d(\Delta\phi)}{|\Delta\textbf{r}|}.
\end{equation}
Next, we apply to (\ref{33}) our $m=0$ integral formula (\ref{30}), perform a Taylor expansion in $a/\mathcal{L}$ and keep terms up to first order, 
\begin{multline}\label{36}
    \textbf{A}_n=\frac{\mu_0I\textbf{e}_1}{4\pi^2a^2}\int_0^{2\pi}d\tilde\theta\int_0^a d\tilde s\tilde s\bigg[-(\kappa s\cos\theta+\kappa \tilde s\cos\tilde\theta)
    \\+\left(1+\frac{\kappa s\cos\theta-\kappa\tilde s\cos\tilde \theta}{2}\right)
    \\\times\left(\ln\left(\frac{2\phi_0^2|\textbf{r}_c'|^2}{s\tilde s}\right)-\ln\left(\frac{s^2+\tilde s^2}{2s\tilde s}-\cos\Delta\theta\right)\right)\bigg].
\end{multline}
Here, we make use of the identities
\begin{subequations}\label{37}
    \begin{gather}
        \int_0^{2\pi}d\tilde\theta\ln(b-\cos \tilde\theta)=-2\pi\ln\left(\frac{2}{b+\sqrt{b^2-1}}\right)\\
        \int_0^{2\pi}d\tilde\theta\cos \tilde\theta\ln(b-\cos \Delta\theta)=2\pi(\sqrt{b^2-1}-b)\cos\theta,
    \end{gather}
\end{subequations}
where $b=(s^2+\tilde s^2)/2s\tilde s$ and $\sqrt{b^2-1}=|s^2-\tilde s^2|/2s\tilde s$. Taking the integral of (\ref{36}) with respect to $\tilde\theta$ then $\tilde s$, we find that
\begin{equation}
    \textbf{A}_n\equiv\begin{cases}
        \textbf{A}_n^<, & s\leq a\\
         \textbf{A}_n^>, & s>a,
    \end{cases}
\end{equation}
where
\begin{subequations}\label{40}
    \begin{gather}
    \begin{split}
            \textbf{A}_n^<=\frac{\mu_0I\textbf{e}_1}{4\pi a^2}\bigg[a^2(1-\kappa s\cos\theta)-s^2\left(1+\frac{\kappa s\cos\theta}{4}\right)
    \\+2a^2\left(1+\frac{\kappa s\cos\theta}{2}\right)\left(\ln\left(\frac{|\textbf{r}_c'|}{a}\right)+\ln(2\phi_0)\right)\bigg]
        \end{split}\\
        \begin{split}
            \textbf{A}_n^>=\frac{\mu_0I\textbf{e}_1}{4\pi a^2}\bigg[-\frac{\kappa a^4\cos\theta}{4s}-\kappa sa^2\cos\theta
            \\+2a^2\left(1+\frac{\kappa s\cos\theta}{2}\right)\left(\ln\left(\frac{|\textbf{r}_c'|}{s}\right)+\ln(2\phi_0)\right)\bigg].
        \end{split} 
    \end{gather}
\end{subequations}
Adding the near and far terms gives the desired result, (\ref{10}).

\section{Reduced Integral Formula for the Self-Inductance\label{appendix:induct}}
The self-inductance of a conductor is given by \cite{jackson1999classical}
\begin{equation}\label{66}
    L=\frac{\mu_0}{4\pi I^2}\int d^3 \textbf{r}\int d^3\tilde {\textbf{r}}\frac{\tilde{\textbf{J}}\cdot\textbf{J}}{|\Delta\textbf{r}|}.
\end{equation}
Applying (\ref{126}), the self-inductance can be written as
\begin{equation}\label{68}
    L=\frac{1}{I^2}\int d^3\textbf{r}(\textbf{J}\cdot\textbf{A}).
\end{equation}
Applying our choice of a uniform current density in the $\textbf{e}_1$ direction (\ref{130}) and taking the integral in the Frenet-Serret coordinate system, (\ref{68}) becomes
\begin{equation}\label{125}
    L=\frac{1}{I\pi a^2}\int_0^{2\pi}d\phi\int_0^{2\pi}d\theta\int_0^a ds\,s(1-\kappa s\cos\theta)|\textbf{r}_c'|(\textbf{e}_1\cdot\textbf{A}).
\end{equation}
As in Appendix \ref{appendix:vecpot}, it is helpful to separate the integrals into near and far regions of the domain,
\begin{equation}\label{70}
    L=L_n+L_f,
\end{equation}
where
\begin{subequations}\label{71}
    \begin{gather}
        L_n=\frac{1}{I\pi a^2}\int_0^{2\pi}\hspace{-0.3em}d\phi\int_0^{2\pi}\hspace{-0.3em}d\theta\int_0^a ds\,s(1-\kappa s\cos\theta)|\textbf{r}_c'|(\textbf{e}_1\cdot\textbf{A}_n)\label{71a}\\
        L_f=\frac{1}{I\pi a^2}\int_0^{2\pi}\hspace{-0.3em}d\phi\int_0^{2\pi}\hspace{-0.3em}d\theta\int_0^a ds\,s(1-\kappa s\cos\theta)|\textbf{r}_c'|(\textbf{e}_1\cdot\textbf{A}_f)\label{71b}.
    \end{gather}
\end{subequations}
We also introduce a regularized self-inductance,
\begin{equation}\label{72}
    L_\textrm{reg}\equiv\frac{\mu_0}{4\pi}
    \int_0^{2\pi}d\phi
    \int_0^{2\pi}d\tilde\phi
    \frac{\textbf{r}_c'\cdot\tilde{\textbf{r}}_c'}{\sqrt{|\Delta\textbf{r}_c|^2+ka^2}},
\end{equation}
where $k$ is some constant.
It will be shown below that $L \approx L_\textrm{reg}$ when $a/\mathcal{L}\ll 1$, for appropriate choice of $k$. 

\subsection{The Regularized Self-Inductance}
As with the full self-inductance, we write the regularized self-inductance as a sum of near and far parts,
\begin{equation}\label{73}
    L_\textrm{reg}=L_{\textrm{reg},n}+L_{\textrm{reg},f},
\end{equation}
where
\begin{subequations}\label{74}
    \begin{gather}
        L_{\textrm{reg},n}=\frac{\mu_0}{4\pi}
        \int_0^{2\pi}d\phi
        \int_{\phi-\phi_0}^{\phi+\phi_0}d\tilde\phi
        \frac{\textbf{r}_c'\cdot\tilde{\textbf{r}}_c'}{\sqrt{|\Delta\textbf{r}_c|^2+ka^2}}\label{74a}\\
        L_{\textrm{reg},f}=\frac{\mu_0}{4\pi}
        \int_0^{2\pi}d\phi
        \int_{\phi+\phi_0}^{2\pi+\phi-\phi_0}d\tilde\phi
        \frac{\textbf{r}_c'\cdot\tilde{\textbf{r}}_c'}{\sqrt{|\Delta\textbf{r}_c|^2+ka^2}}\label{74b}.
    \end{gather}
\end{subequations}
In the far region we may apply the assumption $a/\mathcal{L}\ll 1$ to the denominator of (\ref{74b}), giving
\begin{equation}\label{75}
    L_{\textrm{reg},f}=\frac{\mu_0}{4\pi}
    \int_0^{2\pi}d\phi
    \int_{\phi+\phi_0}^{2\pi+\phi-\phi_0}d\tilde\phi
    \frac{\textbf{r}_c' \cdot \tilde{\textbf{r}}_c'}{|\Delta\textbf{r}_c|}.
\end{equation}
In the near region, we apply (\ref{23}) and write the denominator of (\ref{74a}) as 
\begin{equation}
    |\Delta\textbf{r}_c|^2+ka^2=\alpha+\beta\Delta\phi+\gamma(\Delta\phi)^2,
\end{equation}
where
\begin{subequations}
   \begin{gather}
        \alpha=ka^2\\
    \beta=0\\
    \gamma=|\textbf{r}_c'|^2.
   \end{gather}
\end{subequations}
Note that $\alpha\sim a^2$ while $\gamma\sim \mathcal{L}^2$. Just as in Appendix \ref{appendix:vecpot}, we may express (\ref{74a}) as a sum of integrals of the form (\ref{262}). As we noted previously, because $\alpha/\gamma\ll\phi_0^2$, only the $m=0$ integral meaningfully contributes, as the $m=1$ and $m=2$ integrals are smaller by factors of $(a/\mathcal{L})^2$ and $\phi_0^2$. Therefore, we choose to keep terms only up to zeroth order in $\Delta\phi$ in our expansion of $\tilde{\textbf{r}}_c$. Performing this expansion, (\ref{74a}) becomes
\begin{equation}\label{78}
    L_{\textrm{reg},n}=\frac{\mu_0}{4\pi}\int_0^{2\pi}d\phi|\textbf{r}_c'|^2\int_{\phi-\phi_0}^{\phi+\phi_0}d\tilde\phi\frac{1}{\sqrt{|\Delta\textbf{r}_c|^2+ka^2}}.
\end{equation}
Applying (\ref{28}), where we note that $|\Delta\textbf{r}|\rightarrow\sqrt{|\Delta \textbf{r}_c|^2+ka^2}$ here, (\ref{78}) becomes
\begin{equation}\label{80}
    L_{\textrm{reg},n}=\frac{\mu_0}{4\pi}\int_0^{2\pi}d\phi|\textbf{r}_c'|\left(2\ln\left(\frac{|\textbf{r}_c'|}{a}\right)+ 2\ln(2\phi_0)-\ln k\right).
\end{equation}

\subsection{Far Region}
Applying the far form of the vector potential (\ref{16}) and noting that $|\textbf{r}_c'|\textbf{e}_1=\textbf{r}_c'$, the self-inductance in the far region (\ref{71b}) becomes
\begin{multline}
    L_f=\frac{\mu_0}{4\pi^2a^2 }
    \int_0^{2\pi}d\phi
    %\hspace{-1em}
    \int\limits_{\phi+\phi_0}^{2\pi+\phi-\phi_0}
    %\hspace{-1.5em}
    d\tilde{\phi}
    \int_0^{2\pi}d\theta\int_0^a ds s(1-\kappa s\cos\theta)
    \\\times\frac{\textbf{r}_c'\cdot\tilde{\textbf{r}}_c'}{|\Delta \textbf{r}_c|}\bigg[1-
   \frac{\Delta \textbf{r}_c}{|\Delta \textbf{r}_c|^2}\cdot(s\cos\theta\textbf{e}_2+s\sin\theta\textbf{e}_3)\bigg].
\end{multline}
Keeping only leading-order terms in $a/\mathcal{L}$ and performing the integrals over $\theta$ and $s$, 
\begin{equation}\label{83}
     L_f=\frac{\mu_0}{4\pi}
     %\hspace{-1em}
     \int_0^{2\pi}d\phi
     \int\limits_{\phi+\phi_0}^{2\pi+\phi-\phi_0}
     %\hspace{-1.5em}
     d\tilde{\phi}
     \frac{\textbf{r}_c'\cdot\tilde{\textbf{r}}_c'}{|\Delta \textbf{r}_c|}.
\end{equation}
\subsection{Near Region}
Applying the near form of the vector potential (\ref{40}) and keeping terms up to first order in $a/\mathcal{L}$, the self-inductance in the near region (\ref{71a}) becomes
\begin{multline}
    L_n=\frac{\mu_0}{4\pi^2a^2}\int_0^{2\pi}d\phi\int_0^{2\pi}d\theta\int_0^a ds\,s|\textbf{r}_c'|
    \\\times\bigg[a^2(1-2\kappa s\cos\theta)-s^2\left(1-\frac{3\kappa s\cos\theta}{4}\right)
    \\+2a^2\left(1-\frac{\kappa s\cos\theta}{2}\right)\left(\ln\left(\frac{|\textbf{r}_c'|}{a}\right)+\ln(2\phi_0)\right)\bigg].
\end{multline}
Integrating over $\theta$ and $s$, 
\begin{equation}\label{86}
    L_n=\frac{\mu_0}{2\pi}\int_0^{2\pi}d\phi|\textbf{r}_c'|\left(\frac{1}{4}+\ln(2\phi_0)+\ln\left(\frac{|\textbf{r}_c'|}{a}\right)\right).
\end{equation}
\subsection{The Total Self-Inductance}
As noted in (\ref{70}), the total self-inductance is given by the sum of its near (\ref{86}) and far (\ref{83}) parts. Substituting in the expanded form of the regularized self-inductance which is given as a sum of its own near (\ref{80}) and far (\ref{75}) parts, 
\begin{equation}
    L=L_\textrm{reg}+\frac{\mu_0}{2\pi}\left(\frac{1}{4}+\frac{\ln k}{2}\right)\int_0^{2\pi}d\phi|\textbf{r}_c'|.
\end{equation}
We make the convenient choice that
\begin{equation}\label{89}
    k=\frac{1}{\sqrt{e}}
\end{equation}
such that $L=L_\textrm{reg}$. Thus, we obtain (\ref{91}).

\section{Reduced Integral Formula for the Magnetic Field\label{appendix:mag}}
In this section, we find the simplified magnetic field (\ref{102})-(\ref{63}) as the curl of the vector potential computed in Appendix \ref{appendix:vecpot}.

\subsection{The Regularized Biot-Savart Law}
We wish to find an expanded form of the regularized Biot-Savart law (\ref{43b}),
where in the denominator we have chosen $k=1/\sqrt{e}$ in agreement with (\ref{89}).
To do so, we  write a sum of its near and far region components, 
\begin{equation}\label{reg0}
    \textbf{B}_\textrm{reg}=\textbf{B}_{\textrm{reg},n}+\textbf{B}_{\textrm{reg},f},
\end{equation}
where
\begin{subequations}\label{reg1}
    \begin{gather}
        \textbf{B}_{\textrm{reg},n}=\frac{\mu_0I}{4\pi}\int_{\phi-\phi_0}^{\phi+\phi_0}d\tilde\phi\frac{\tilde{\textbf{r}}_c'\times\Delta\textbf{r}_c}{\left(|\Delta\textbf{r}_c|^2+\frac{a^2}{\sqrt{e}}\right)^\frac{3}{2}}\label{reg1a}\\
        \textbf{B}_{\textrm{reg},f}=\frac{\mu_0I}{4\pi}\int_{\phi+\phi_0}^{2\pi+\phi-\phi_0}d\tilde\phi\frac{\tilde{\textbf{r}}_c'\times\Delta\textbf{r}_c}{\left(|\Delta\textbf{r}_c|^2+\frac{a^2}{\sqrt{e}}\right)^\frac{3}{2}}.\label{reg1b}
    \end{gather}
\end{subequations}
The far region integral (\ref{reg1b}) can be simplified for $a\ll|\Delta \textbf{r}|$,
\begin{equation}\label{reg2}
    \textbf{B}_{\textrm{reg},f}\approx\frac{\mu_0I}{4\pi}\int_{\phi+\phi_0}^{2\pi+\phi-\phi_0}d\tilde\phi\frac{\tilde{\textbf{r}}_c'\times\Delta\textbf{r}_c}{|\Delta\textbf{r}_c|^\frac{3}{2}}.
\end{equation}
As in Appendix \ref{appendix:vecpot}, the near region integral can be simplified by applying $|\Delta\phi|\ll 1$ to the numerator and denominator of (\ref{reg1a}). Noting that $\tilde{\textbf{r}}_c'=|\tilde{\textbf{r}}_c'|\tilde{\textbf{e}}_1$, we use (\ref{21}) to write that
\begin{equation}\label{reg4}
\tilde{\textbf{r}}_c'\times\Delta\textbf{r}_c=\frac{\kappa|\textbf{r}_c'|^3(\Delta\phi)^2\textbf{e}_3}{2},
\end{equation}
where we Taylor expanded $\tilde\kappa$ and $|\tilde{\textbf{r}}_c'|$ and kept terms up to second order in $\Delta\phi$. Applying this result (\ref{reg4}) and $|\Delta\textbf{r}_c|^2=|\textbf{r}_c'|^2(\Delta\phi)^2$, (\ref{reg1a}) becomes
\begin{equation}\label{reg6}
    \textbf{B}_{\textrm{reg},n}=\frac{\mu_0I\kappa\textbf{e}_3}{8\pi}\int_{-\phi_0}^{\phi_0}d(\Delta\phi)\frac{(\Delta\phi)^2}{\left((\Delta\phi)^2+\frac{a^2}{\sqrt{e}|\textbf{r}_c'|^2}\right)^\frac{3}{2}}.
\end{equation}
We apply\cite{gradshteyn2014table}
\begin{multline}\label{reg7}
    \int_{-\phi_0}^{\phi_0}d(\Delta\phi)\frac{(\Delta\phi)^2}{((\Delta\phi)+\alpha)^\frac{3}{2}}
    \\=\ln\alpha-2\phi_0\sqrt{\frac{1}{\alpha+\phi_0^2}}-2\ln\left(-\phi_0+\sqrt{\alpha+\phi_0^2}\right),
\end{multline}
where we identify $\alpha=a^2/\sqrt{e}|\textbf{r}_c'|^2$ in (\ref{reg6}). From our standard assumption that $a/\mathcal{L}\ll\phi_0$, we have that $\alpha\ll\phi_0^2$. Under this limit, (\ref{reg7}) simplifies to
\begin{equation}
    \int_{-\phi_0}^{\phi_0}d(\Delta\phi)\frac{(\Delta\phi)^2}{((\Delta\phi)+\alpha)^\frac{3}{2}}\approx 2\ln (2\phi_0)-\ln\alpha-2.
\end{equation}
Applying this result to (\ref{reg6}), we have
\begin{equation}\label{reg9}
    \textbf{B}_{\textrm{reg},n}=\frac{\mu_0I\kappa\textbf{e}_3}{8\pi}\left(2\ln (2\phi_0)+2\ln\left(\frac{|\textbf{r}_c'|}{a}\right)-\frac{3}{2}\right).
\end{equation}
\subsection{Far Region}
Here we wish to evaluate $\nabla\times\textbf{A}_f$, where the vector potential in the far region is given by  (\ref{162}). The curl operator acts both on the integrand and the integration limits in accordance with the Leibniz rule,

\begin{equation}\label{45}
    \frac{d}{dx}\hspace{-0.25em}\int\limits_{a(x)}^{b(x)}\hspace{-0.25em}f(x,t)dt=\hspace{-0.25em}\int\limits_{a(x)}^{b(x)}\hspace{-0.25em}dt\frac{\partial f}{\partial x}+f(x,b(x))\frac{db}{dx}-f(x,a(x))\frac{da}{dx}.
\end{equation}
We may use the formula for curl in non-orthogonal coordinate systems \cite{khare1970divergence},
  \begin{equation}\label{8}
      \nabla\times\textbf{C}=\frac{1}{\sqrt{g}}\begin{vmatrix}
\partial\textbf{r}/\partial s & \partial\textbf{r}/\partial\theta & \partial\textbf{r}/\partial\phi\\
\partial/\partial s & \partial/\partial \theta & \partial/\partial \phi \\
\textbf{C}\cdot\partial\textbf{r}/\partial s & \textbf{C}\cdot\partial\textbf{r}/\partial \theta & \textbf{C}\cdot\partial\textbf{r}/\partial \phi
\end{vmatrix},
  \end{equation}
where the derivatives of the position vector can be obtained from (\ref{4}). Applying (\ref{45}),
\begin{multline}\label{46}
    \nabla\times\int_{a(\phi)}^{b(\phi)}\textbf{f}(\textbf{r},\phi')d\phi'=\int_{a(\phi)}^{b(\phi)}\nabla\times\textbf{f}(\textbf{r},\phi')d\phi'
    \\+\frac{1}{\sqrt{g}}\bigg[\frac{\partial\textbf{r}}{\partial\theta}\left(\left(\frac{db}{d\phi}\textbf{f}(\textbf{r},b(\phi))-\frac{da}{d\phi}\textbf{f}(\textbf{r},a(\phi))\right)\cdot\frac{\partial\textbf{r}}{\partial s}\right)
    \\-\frac{\partial\textbf{r}}{\partial s}\left(\left(\frac{db}{d\phi}\textbf{f}(\textbf{r},b(\phi))-\frac{da}{d\phi}\textbf{f}(\textbf{r},a(\phi))\right)\cdot\frac{\partial\textbf{r}}{\partial \theta}\right)\bigg].
\end{multline}
Let us denote the curl of (\ref{162}) as 
\begin{equation}\label{47}
    \nabla\times\textbf{A}_f=
[\nabla\times\textbf{A}_f]_\textrm{int}+[\nabla\times\textbf{A}_f]_\textrm{dom},
\end{equation}
where $[\nabla\times\textbf{A}_f]_\textrm{int}$ refers to the differentiation of the integrand as in the first line of (\ref{46}) and $[\nabla\times\textbf{A}_f]_\textrm{dom}$ refers to the differentiation of the domain as in the last two lines of (\ref{46}). 

Let us start with finding $[\nabla\times\textbf{A}_f]_\textrm{int}$, which is given by
\begin{equation}
    [\nabla\times\textbf{A}_f]_\textrm{int}=\frac{\mu_0I}{4\pi}\hspace{-2em}\int\limits_{\hspace{1em}\phi+\phi_0}^{\hspace{1em}2\pi+\phi-\phi_0}\hspace{-1.5em}d\tilde{\phi}\left(\nabla\times\frac{\tilde{\textbf{r}}_c'}{|\textbf{r}-\tilde{\textbf{r}}_c|}\right).
\end{equation}
Applying the product rule for a curl and taking the gradient in Cartesian coordinates, we find
\begin{equation}
    [\nabla\times\textbf{A}_f]_\textrm{int}=\frac{\mu_0I}{4\pi}\hspace{-2em}\int\limits_{\hspace{1em}\phi+\phi_0}^{\hspace{1em}2\pi+\phi-\phi_0}\hspace{-1em}d\tilde{\phi}\frac{\tilde{\textbf{r}}_c'\times(\textbf{r}-\tilde{\textbf{r}}_c)}{|\textbf{r}-\tilde{\textbf{r}}_c|^3}.
\end{equation}
Using the definition of the position vector (\ref{4}) and the far region assumption,  $a\ll|\Delta\textbf{r}|$, this simplifies to
\begin{equation}\label{51}
    [\nabla\times\textbf{A}_f]_\textrm{int}=\frac{\mu_0I}{4\pi}\hspace{-2em}\int\limits_{\hspace{1em}\phi+\phi_0}^{\hspace{1em}2\pi+\phi-\phi_0}\hspace{-1em}d\tilde{\phi}\frac{\tilde{\textbf{r}}_c'\times(\textbf{r}-\tilde{\textbf{r}}_c)}{|\Delta\textbf{r}_c|^3}.
\end{equation}

Next, let us consider $[\nabla\times\textbf{A}_f]_\textrm{dom}$. Applying the formulas for the position vector (\ref{4}) and the Jacobian (\ref{7}), we can write with some algebra that
\begin{equation}\label{52}
     [\nabla\times\textbf{A}_f]_\textrm{dom}=\frac{\mu_0I}{4\pi}\frac{\textbf{e}_3\textbf{e}_2-\textbf{e}_2\textbf{e}_3}{(1-\kappa s\cos\theta)|\textbf{r}_c'|}\cdot\left[\frac{\tilde{\textbf{r}}_c'}{|\textbf{r}-\tilde{\textbf{r}}_c|}\right]_{\phi+\phi_0}^{\phi-\phi_0}. 
\end{equation}
In order to simplify this expression, let us expand $\tilde{\textbf{r}}_c'$ and $|\textbf{r}-\tilde{\textbf{r}}_c|^{-1}$ about $\tilde\phi=\phi$ and keep terms up to $\Delta\phi$. Noting that $\tilde{\textbf{r}}_c'=|\tilde{\textbf{r}}_c'|\tilde{\textbf{e}}_1$ and applying the Frenet-Serret formula (\ref{2}), 
\begin{equation}\label{53}
    \tilde{\textbf{r}}_c'\approx|{\textbf{r}}_c'|\textbf{e}_1-\Delta\phi\left(\frac{d|\textbf{r}_c'|}{d\phi}\textbf{e}_1+\kappa|\textbf{r}_c'|^2\textbf{e}_2\right).
\end{equation}
We next perform a Taylor series expansion of $|\Delta\textbf{r}_c|^{-1}$, 
\begin{equation}\label{55}
    \frac{1}{|\Delta\textbf{r}_c|}\approx \frac{1}{|\Delta\phi||\tilde{\textbf{r}}_c'|}\left(1+\frac{\Delta\phi}{2|\tilde{\textbf{r}}_c'|}\frac{d|\tilde{\textbf{r}}_c'|}{d\phi}\right),
\end{equation}
where $|\Delta\textbf{r}_c|$ can be found from (\ref{19}). Combining (\ref{53}) and (\ref{55}), 
\begin{equation}
    \left[\frac{\tilde{\textbf{r}}_c'}{|\textbf{r}-\tilde{\textbf{r}}_c|}\right]_{\phi+\phi_0}^{\phi-\phi_0}=-\left(\frac{1}{|\tilde{\textbf{r}}_c'|}\frac{d|\tilde{\textbf{r}}_c'|}{d\phi}\textbf{e}_1+2\kappa|\tilde{\textbf{r}}_c'|\textbf{e}_2\right).
\end{equation}
Substituting this result into (\ref{52}) and keeping terms up to first order in $a/\mathcal{L}$, we find that
\begin{equation}\label{57}
     [\nabla\times\textbf{A}_f]_\textrm{dom}=-\frac{\mu_0I\kappa\textbf{e}_3}{2\pi}.
\end{equation}

At last, in accordance with (\ref{47}), we sum the two contributions to the curl, (\ref{51}) and (\ref{57}), to find the total curl in the far region,
\begin{equation}\label{58}
    \nabla\times\textbf{A}_f=\frac{\mu_0I}{4\pi}\hspace{-2em}\int\limits_{\hspace{1em}\phi+\phi_0}^{\hspace{1em}2\pi+\phi-\phi_0}\hspace{-1em}d\tilde{\phi}\frac{\tilde{\textbf{r}}_c'\times(\textbf{r}-\tilde{\textbf{r}}_c)}{|\textbf{r}-\tilde{\textbf{r}}_c'|^3}-\frac{\mu_0I\kappa\textbf{e}_3}{2\pi}.
\end{equation}
\subsection{Near Region}
Here we wish to evaluate $\nabla\times\textbf{A}_n$, where the vector potential in the near region is specified in (\ref{40}). Applying  (\ref{8}), performing the expansion $(1-\kappa s\cos\theta)^{-1}\approx 1+\kappa s\cos\theta$, and keeping terms up to first order in $a/\mathcal{L}$, we find
\begin{subequations}\label{61}
    \begin{align}
        \begin{split}
            \nabla\times\textbf{A}_n^<&=\frac{\mu_0I}{2\pi a^2}\bigg[\left(s\cos\theta+\kappa s^2\left(\frac{\cos(2\theta)}{8}-\frac{1}{4}\right)
    \right.\\& \left.+\frac{\kappa a^2}{2}\left(2+\ln\left(\frac{|\textbf{r}_c'|}{a}\right)+\ln (2\phi_0)\right)\right)\textbf{e}_3
    \\&+\left(-s\sin\theta-\frac{\kappa s^2\sin(2\theta)}{8}\right)\textbf{e}_2\bigg].
        \end{split}\\
        \begin{split}
            \nabla\times\textbf{A}_n^>&=\frac{\mu_0I}{2\pi a^2}\bigg[\left(\frac{a^2\cos\theta}{s}-\frac{\kappa a^4\cos (2\theta)}{8s^2}+\frac{\kappa a^2\cos (2\theta)}{4}
    \right.\\& \left.+\frac{\kappa a^2}{2}\left(\frac{3}{2}+\ln\left(\frac{|\textbf{r}_c'|}{s}\right)+\ln (2\phi_0)\right)\right)\textbf{e}_3
    \\&+\left(-\frac{a^2\sin\theta}{s}+\frac{\kappa a^4\sin (2\theta)}{8s^2}-\frac{\kappa a^2\sin (2\theta)}{4}\right)\textbf{e}_2
    \bigg].
        \end{split}
    \end{align}
\end{subequations}

\subsection{The Total Magnetic Field}
As stated in (\ref{11}), the total magnetic vector potential is given by the sum of its near and far parts. As a result, the total magnetic field, taken as $\textbf{B}=\nabla\times\textbf{A}$, is given as the sum of curls of the far part (\ref{58}) and the near part (\ref{61}),
\begin{multline}\label{129}
    \textbf{B}=\frac{\mu_0I}{4\pi}\int_{\phi+\phi_0}^{2\pi+\phi-\phi_0}d\phi'\frac{\tilde{\textbf{r}}_c'\times\Delta\textbf{r}_c}{|\Delta\textbf{r}_c|^\frac{3}{2}}
    \\+\frac{\mu_0I}{2\pi a^2}\begin{cases}
        \left(-s\sin\theta-\frac{\kappa s^2\sin(2\theta)}{8}\right)\textbf{e}_2
    \\+\left(s\cos\theta+\kappa s^2\left(\frac{\cos(2\theta)}{8}-\frac{1}{4}\right)
    \right. & s\leq a\\  \left.+\frac{\kappa a^2}{2}\left(\ln\left(\frac{|\textbf{r}_c'|}{a}\right)+\ln(2\phi_0)\right)\right)\textbf{e}_3 
    \vspace{1em}\\
    \left(-\frac{a^2\sin\theta}{s}+\frac{\kappa a^4\sin (2\theta)}{8s^2}-\frac{\kappa a^2\sin (2\theta)}{4}\right)\textbf{e}_2
    \\+\left(\frac{a^2\cos\theta}{s}-\frac{\kappa a^4\cos (2\theta)}{8s^2}+\frac{\kappa a^2\cos (2\theta)}{4}
    \right. & s>a.\\ \left. +\frac{\kappa a^2}{2}\left(-\frac{1}{2}+\ln\left(\frac{|\textbf{r}_c'|}{s}\right)+\ln (2\phi_0)\right)\right)\textbf{e}_3
    \end{cases}
\end{multline}
Comparing this expression to the expanded form of the regluarized Biot-Savart law, which is given as the sum (\ref{reg0}) of its far (\ref{reg2}) and near (\ref{reg9}) parts, we write the magnetic field as (\ref{102}).

\section{Reduced Integral Formula for the Self-Force\label{appendix:force}}
The self-force per unit length (abbreviated here as self-force) can be calculated in at least two ways. An elegant approach is to use the principle of virtual work, examined in a similar context in section 3 of \cite{garren1994lorentz}, in which the shape of the coil is perturbed and the change to the configuration's energy is examined. Expressing the latter in terms of the inductance, and applying (\ref{91}), one obtains the self-force (\ref{103}) with (\ref{43b}). Alternatively, we can directly use the formula for total force as a volume integral of $\textbf{J}\times\textbf{B}$,
\begin{equation}\label{104}
\textbf{F}=\frac{I}{\pi a^2}\int_0^{2\pi}d\phi\int_0^{2\pi}d\theta\int_0^ads s(1-\kappa s\cos\theta)|\textbf{r}_c'|(\textbf{e}_1\times\textbf{B}),
\end{equation}
where we have applied our choice of uniform current density in the $\textbf{e}_1$ direction. Comparing this expression to the total force as a line integral of the force per unit length,
\begin{equation}
\textbf{F}=\int_0^{2\pi}d\phi|\textbf{r}_c'|\frac{d\textbf{F}}{ dl},
\end{equation}
the self-force must be given by
\begin{equation}\label{123}
\frac{d\textbf{F}}{dl}=\frac{I}{\pi a^2}\int_0^{2\pi} d\theta\int_0^a ds s(1-\kappa s\cos\theta)(\textbf{e}_1\times\textbf{B}).
\end{equation}
Next, we substitute in the one-dimensional integral formula for the magnetic field (\ref{102})-(\ref{63}) and keep terms up to first order in $a/\mathcal{L}$,
\begin{multline}
    \frac{d\textbf{F}}{dl}=I\textbf{e}_1\times\textbf{B}_\textrm{reg}+\frac{\mu_0I^2}{4\pi^2a^2}\int_0^{2\pi}d\theta\int_0^a ds s\\\times\bigg[\bigg(-\frac{2s\cos\theta}{a^2}
    +\frac{2\kappa s^2\cos^2\theta}{a^2}-\frac{3\kappa}{4}+\frac{\kappa s^2}{2a^2}-\frac{\kappa s^2}{4a^2}\cos(2\theta)\bigg){\textbf{e}}_2\\
    +\left(-\frac{2s}{a^2}\sin\theta+\frac{3\kappa s^2}{4a^2}\sin(2\theta)\right){\textbf{e}}_3\bigg],
\end{multline}
where the regularized Biot-Savart law is given in (\ref{43b}). Performing the integrals over $\theta$ and $s$, we find the force to be (\ref{103}).

\section*{Acknowledgment}

We thank 
Robert Granetz,
Stuart Hudson,
Thomas Kruger, 
Jorrit Lion,
Nicolo Riva,
Felix Warmer,
Yuhu Zhai, and
Caoxiang Zhu
for useful discussions.
This work was supported by a grant from the Simons Foundation (No. 560651, T. A.).
This work was also supported by the U.S. Department of Energy under Contract DE-FG02-93ER54197.

\ifCLASSOPTIONcaptionsoff
  \newpage
\fi



\bibliographystyle{IEEEtran}
\bibliography{main}
\end{document}